\documentclass[twocolumn,usenatbib,floatfix]{aastex62}

\usepackage{graphicx}
\usepackage{float}

\newcommand{\asec}{$^{\prime \prime} \,$}
\graphicspath{{./}{figures/}}

\received{...}
\revised{...}
\accepted{...}
\submitjournal{ApJ}

\shorttitle{}
\shortauthors{A. Bonafede et al.}

\begin{document}

\title{The Coma cluster at LOFAR frequencies II:  the halo, relic, and a new accretion relic}

\correspondingauthor{Annalisa Bonafede et al}
\email{annalisa.bonafede@unibo.it}

\author{A. Bonafede}
\affiliation{DIFA - Universit\`a di Bologna, via Gobetti 93/2, I-40129 Bologna, Italy }
\affiliation{INAF - IRA, Via Gobetti 101, I-40129 Bologna, Italy; IRA - INAF, via P. Gobetti 101, I-40129 Bologna, Italy;} 
\affiliation{Universit\"at Hamburg, Hamburger Sternwarte, Gojenbergsweg 112, 21029, Hamburg, Germany; }

\author{G. Brunetti}
\affiliation{INAF - IRA, Via Gobetti 101, I-40129 Bologna, Italy; IRA - INAF, via P. Gobetti 101, I-40129 Bologna, Italy;}

\author{L. Rudnick}
\affiliation{Minnesota Institute for Astrophysics, School of Physics and Astronomy, University of Minnesota, Minneapolis, MN 55455, USA}

\author{F. Vazza}
\affiliation{DIFA - Universit\`a di Bologna, via Gobetti 93/2, I-40129 Bologna, Italy }
\affiliation{INAF - IRA, Via Gobetti 101, I-40129 Bologna, Italy; IRA - INAF, via P. Gobetti 101, I-40129 Bologna, Italy;}
\affiliation{Universit\"at Hamburg, Hamburger Sternwarte, Gojenbergsweg 112, 21029, Hamburg, Germany; }

\author{H. Bourdin} 
\affiliation{Universit\`a di Roma Tor Vergata, Via della Ricerca Scientifica, I00133 Roma, Italy;}
\affiliation{INFN, Sezione di Roma 2, Universit\`a di Roma Tor Vergata, Via della Ricerca Scientifica, 1, Roma, Italy;}

\author{G. Giovannini}
\affiliation{DIFA - Universit\`a di Bologna, via Gobetti 93/2, I-40129 Bologna, Italy }
\affiliation{INAF - IRA, Via Gobetti 101, I-40129 Bologna, Italy; IRA - INAF, via P. Gobetti 101, I-40129 Bologna, Italy;} 

\author{T. W. Shimwell}
\affiliation{ASTRON, Netherlands Institute for Radio Astronomy, Oude Hoogeveensedijk 4, 7991 PD, Dwingeloo, The Netherlands;}
\affiliation{Leiden Observatory, Leiden University, PO Box 9513, 2300 RA Leiden, The Netherlands;}

\author{X. Zhang}
\affiliation{Leiden Observatory, Leiden University, PO Box 9513, 2300 RA Leiden, The Netherlands;} \affiliation{SRON Netherlands Institute for Space Research, Niels Bohrweg 4, 2333 CA Leiden, The Netherlands }

\author{P. Mazzotta}
\affiliation{Universit\`a di Roma Tor Vergata, Via della Ricerca Scientifica, I00133 Roma, Italy;}
\affiliation{INFN, Sezione di Roma 2, Universit\`a di Roma Tor Vergata, Via della Ricerca Scientifica, 1, Roma, Italy;}

\author{A. Simionescu}
\affiliation{SRON Netherlands Institute for Space Research, Niels Bohrweg 4, 2333 CA Leiden, The Netherlands }
\affiliation{Leiden Observatory, Leiden University, PO Box 9513, 2300 RA Leiden, The Netherlands;}
\affiliation{Kavli Institute for the Physics and Mathematics of the Universe (WPI), The University of Tokyo, Kashiwa, Chiba 277-8583, Japan; }

\author{N. Biava}
\affiliation{DIFA - Universit\`a di Bologna, via Gobetti 93/2, I-40129 Bologna, Italy }
\affiliation{INAF - IRA, Via Gobetti 101, I-40129 Bologna, Italy; IRA - INAF, via P. Gobetti 101, I-40129 Bologna, Italy;}

\author{E. Bonnassieux}
\affiliation{DIFA - Universit\`a di Bologna, via Gobetti 93/2, I-40129 Bologna, Italy }
\affiliation{INAF - IRA, Via Gobetti 101, I-40129 Bologna, Italy; IRA - INAF, via P. Gobetti 101, I-40129 Bologna, Italy;}

\author{M. Brienza}
\affiliation{DIFA - Universit\`a di Bologna, via Gobetti 93/2, I-40129 Bologna, Italy }
\affiliation{INAF - IRA, Via Gobetti 101, I-40129 Bologna, Italy; IRA - INAF, via P. Gobetti 101, I-40129 Bologna, Italy;}

\author{M. Br\"uggen}
\affiliation{Universit\"at Hamburg, Hamburger Sternwarte, Gojenbergsweg 112, 21029, Hamburg, Germany; }

\author{K. Rajpurohit}
\affiliation{DIFA - Universit\`a di Bologna, via Gobetti 93/2, I-40129 Bologna, Italy }
\affiliation{INAF - IRA, Via Gobetti 101, I-40129 Bologna, Italy; IRA - INAF, via P. Gobetti 101, I-40129 Bologna, Italy;}
\affiliation{Th\"uringer Landessternwarte, Sternwarte 5, 07778 Tautenburg, Germany}

\author{C. J. Riseley}
\affiliation{DIFA - Universit\`a di Bologna, via Gobetti 93/2, I-40129 Bologna, Italy }
\affiliation{INAF - IRA, Via Gobetti 101, I-40129 Bologna, Italy; IRA - INAF, via P. Gobetti 101, I-40129 Bologna, Italy;}
\affiliation{CSIRO Space \& Astronomy, PO Box 1130, Bentley, WA 6102, Australia}

\author{C. Stuardi}
\affiliation{DIFA - Universit\`a di Bologna, via Gobetti 93/2, I-40129 Bologna, Italy }
\affiliation{INAF - IRA, Via Gobetti 101, I-40129 Bologna, Italy; IRA - INAF, via P. Gobetti 101, I-40129 Bologna, Italy;}

\author{L. Feretti}
\affiliation{INAF - IRA, Via Gobetti 101, I-40129 Bologna, Italy; IRA - INAF, via P. Gobetti 101, I-40129 Bologna, Italy;}

\author{C. Tasse}
\affiliation{GEPI and USN, Observatoire de Paris, Universit\'e PSL, CNRS, 5 Place Jules Janssen, 92190 Meudon, France}
\affiliation{Department of Physics and Electronics, Rhodes University, PO Box 94, Grahamstown, 6140, South Africa}

\author{A. Botteon}
\affiliation{Leiden Observatory, Leiden University, PO Box 9513, 2300 RA Leiden, The Netherlands;}

\author{E. Carretti}
\affiliation{INAF - IRA, Via Gobetti 101, I-40129 Bologna, Italy; IRA - INAF, via P. Gobetti 101, I-40129 Bologna, Italy;}

\author{R. Cassano}
\affiliation{INAF - IRA, Via Gobetti 101, I-40129 Bologna, Italy; IRA - INAF, via P. Gobetti 101, I-40129 Bologna, Italy;}

\author{V. Cuciti}
\affiliation{Universit\"at Hamburg, Hamburger Sternwarte, Gojenbergsweg 112, 21029, Hamburg, Germany; }

\author{F. de Gasperin}
\affiliation{Universit\"at Hamburg, Hamburger Sternwarte, Gojenbergsweg 112, 21029, Hamburg, Germany; }
\affiliation{INAF - IRA, Via Gobetti 101, I-40129 Bologna, Italy; IRA - INAF, via P. Gobetti 101, I-40129 Bologna, Italy;}

\author{F. Gastaldello}
\affiliation{INAF - IASF Milano, via A. Corti 12, 20133 Milano, Italy;}

\author{M. Rossetti}
\affiliation{INAF - IASF Milano, via A. Corti 12, 20133 Milano, Italy;}

\author{H. J. A. Rottgering}
\affiliation{Leiden Observatory, Leiden University, PO Box 9513, 2300 RA Leiden, The Netherlands;}

\author{T. Venturi}
\affiliation{INAF - IRA, Via Gobetti 101, I-40129 Bologna, Italy; IRA - INAF, via P. Gobetti 101, I-40129 Bologna, Italy;}

\author {R. J. van Weeren}
\affiliation{Leiden Observatory, Leiden University, PO Box 9513, 2300 RA Leiden, The Netherlands;}

\begin{abstract}
We present LOw Frequency ARray observations of the Coma cluster field at 144\,MHz. 
The cluster hosts one of the most famous radio halos, a relic, and a low surface-brightness bridge. 
We detect new features that allow us to make a step forward in the understanding of particle acceleration in clusters. 
The radio halo extends for more than 2 Mpc, which is the largest extent ever reported. To the North-East of the cluster, 
beyond the Coma virial radius, we discover an arc-like radio source that could trace particles accelerated by an accretion shock. To the West of the halo, coincident with a shock detected in the X-rays, we confirm the presence of a radio front, with different spectral properties with respect to the rest of the halo. We detect a radial steepening of the radio halo spectral index between 144 MHz and 342 MHz, at $\sim 30^{\prime}$ from the cluster centre, that may indicate a non constant re-acceleration time throughout the volume. We also detect a mild steepening of the spectral index towards the cluster centre. For the first time, a radial change in the slope of the radio-X-ray correlation is found, and we show that such a change could indicate an increasing fraction of cosmic ray versus thermal energy density in the cluster outskirts. Finally, we investigate the origin of the emission between the relic and the source NGC~4789, and we argue that NGC4789 could have crossed the shock originating the radio emission visible between its tail and the relic.

\end{abstract}

\keywords{galaxy clusters; non-thermal emission; particle acceleration; radio observations}


\section{Introduction}
\label{sec:intro}

\begin{figure*}
\includegraphics[width=2\columnwidth]{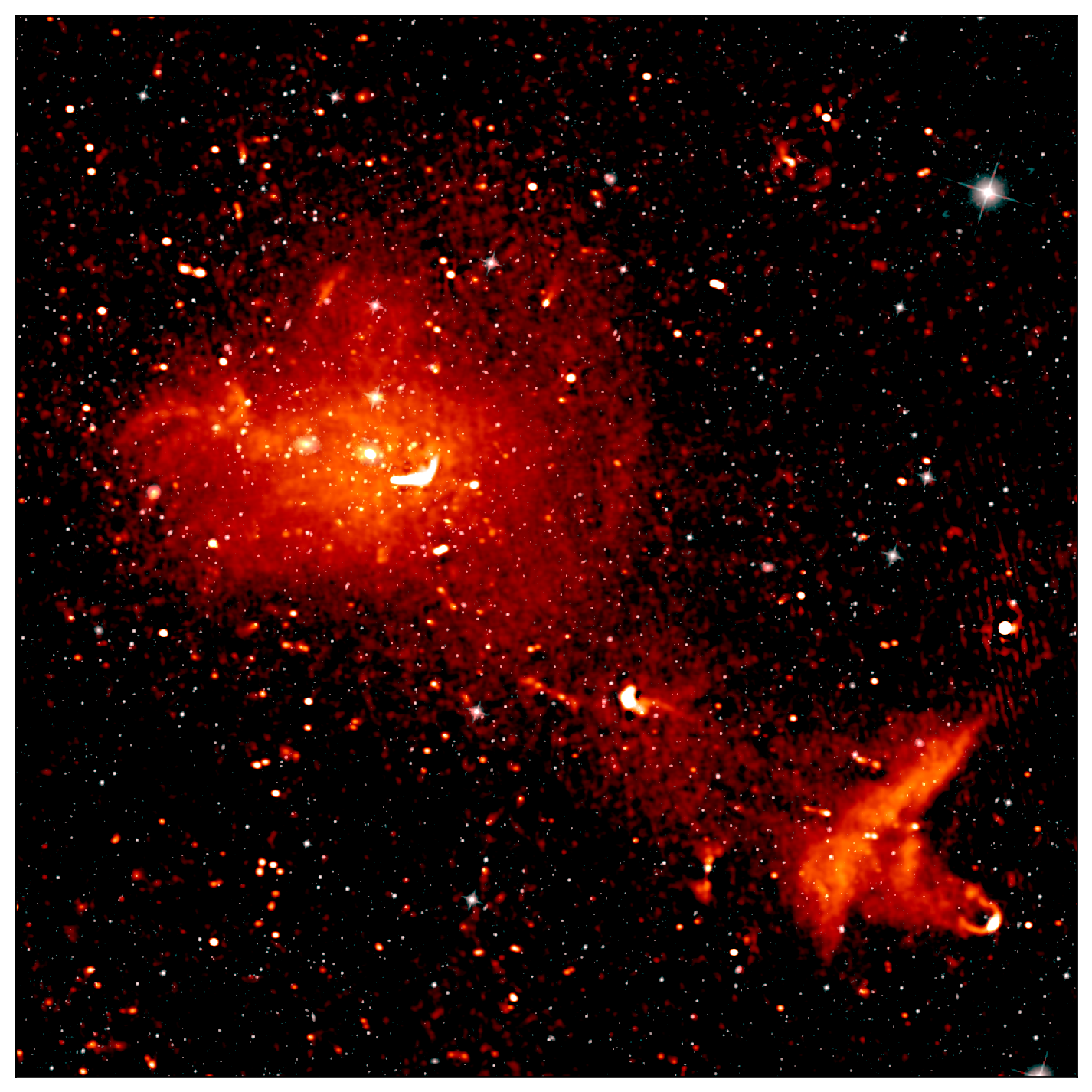}
\caption{Composite IR-radio image of the Coma cluster field. In white the IR image in band 1, 2, and 3 of WISE (\emph{Wide-field Infrared Survey Explorer}) are shown. The red-orange color scale shows the composite radio image of the diffuse emission at 1$^{\prime}$ for the diffuse emission and of 20\asec  for the sources in the field.}
\label{fig:ComaRGB}
\end{figure*}

Diffuse, non-thermal emission has been observed in more than 100 clusters of galaxies, revealing the existence of magnetic fields and relativistic particles on scales as large as a few megaparsecs. In the last decade, the advent of low-frequency, sensitive radio observations has brought about a major advance in the discovery of these objects and the characterization of their properties.\\
Synchrotron emission from the intra-cluster medium (ICM) has been observed in the form of giant radio halos, mini halos, and radio relics, depending on their location and morphology. Giant radio halos are found at the centres of merging galaxy clusters, co-spatial with the X-ray emitting gas, and with sizes of 1-2\,Mpc. Mini halos only have sizes of few hundreds of kpc and are found mostly in cool-core clusters. Finally, radio relics are arc-like sources located in cluster outskirts where they trace merger shock waves. We refer to \cite{vanWeeren19} for a recent review.\\
Since their discovery, it has been proposed that radio halos and relics are generated by shocks and turbulence driven in the ICM by cluster mergers (see e.g. \citealt{BJ14} for a review, and ref. therein). Although the details of the proposed mechanisms are not understood yet, radio halos have preferentially been found in merging clusters, supporting a connection between mergers and radio emission \citep{Cuciti15}, and radio relics are often coincident with gas discontinuities detected in the X-ray and Sunyaev-Zeldovich' (SZ) images \citep[e.g.][]{PlanckComa,OgreanToothbrish}. The origin of mini halos is still debated. They could either originate from 
turbulent motions that develop in the cluster core \citep[e.g.][]{ZuHone13} or from hadronic collisions between cosmic ray protons and thermal protons \citep[e.g][]{Pfrommer04}.\\
In the last years, the picture has become more complicated, and it has been found that some clusters with mini halos also host a larger-scale radio component, resembling a dimmer version of the giant radio halos but in non-merging objects \citep[e.g.,][]{Savini_2019,Raja20}. Moreover, a giant halo has also been found in a strong cool-core cluster CL1821+643 \citep{Bonafede14b}. In addition, relics and halos are sometimes connected through low-brightness radio bridges \citep{vanweeren12,Bonafede21}, that could be powered by mechanisms similar to those that are currently used to explain radio halos. 
On even larger scales, giant bridges of radio emission have been discovered, connecting massive clusters in a pre-merging state \citep[e.g.,][]{Govoni19,Botteon20bridge}.\\
Despite the differences between these types of sources, they all have a low surface brightness ($\sim 1 \mu \rm{Jy/arcsec^2}$) at GHz frequencies. Also they have steep radio spectra \footnote{Throughout this paper, we define the spectral index $\alpha$ as $S(\nu) \propto \nu^{\alpha}$, where $S$ is the flux density at the frequency $\nu$}, with a spectral index $\alpha < -1$ that make them brighter at low radio frequencies. 
Hence, the advent of deep, low-frequency radio surveys, such as the LOFAR Two-metre Sky Survey \citep[LoTSS;][]{Shimwell17,Shimwell19} has both increased the number of new detections (see e.g. \citealt{Biava21b,Riseley21}; Botteon et al., accepted; Hoang et al, submitted) and allowed the study of known objects with unprecedented sensitivity and detail.\\
The Coma cluster hosts the most famous and best studied radio halo, as well as a radio relic and a radio bridge connecting the two (see Fig. \ref{fig:ComaRGB}). The emission from the Coma field has been the subject of many studies since its discovery \citep[e.g.][]{Large59, Ballarati81, Giovannini91, Venturi90, Kronberg07,BrownRudnick11}. 
In paper I \citep{Bonafede21}, we have analysed the properties of the radio bridge and 
we have shown under which conditions it can be powered by turbulent acceleration.
In this paper, we focus on the radio halo and relic of the Coma cluster. New LOFAR data give us information in regions that have been so far inaccessible, providing important clues on the origin of the radio emission.\\
 \cite{Giovannini93} first found that the spectrum of the halo between 326 MHz and 1.28 GHz  is characterized by two different regions: a central one with $\alpha \sim -0.8$ and a peripheral one with $\alpha \sim -1.2$. The integrated spectrum, computed between 30 MHz and 4.8 GHz shows a high-frequency steepening consistent with homogeneous in-situ re-acceleration models. 
 They also reported a smooth distribution of the radio surface brightness, with no evidence for substructures at the resolution of $\sim$50\asec.
 Their results have been confirmed later on by several authors \citep[e.g][]{Thierbach03} who complemented the analysis with observations up to 4.8 GHz, where the halo is barely detected.
 The steepening of the halo at high frequencies has been subject to debate, as the decrement due to the SZ effect was initially not accounted for. After the observation of the SZ effect by the Planck satellite \citep{Planck11,Planck14} it has been possible to confirm the steepening of the radio emission \citep{Brunetti13}. 
However, we note that data below 300 MHz, being taken in the 80s, lack the sensitivity, resolution, calibration and imaging accuracy that are allowed by present instruments and techniques. In particular, the observation at 151 MHz \citep{Cordey85}, i.e. the closest in frequency to LOFAR HBA, has a resolution of $\sim$70\asec, and a sensitivity of a few 10 mJy/beam, that allowed the detection of the central part of the halo only. 
Similarly, observations at 43 and 73 MHz by \cite{HanischErickson} did not allow to properly subtract the emission from radiogalaxies present in the cluster. \\
 In addition to integrated spectral studies, better resolved spatial analysis of radio halos spectral properties yields important information about the distribution of the non-thermal component in the ICM \citep[e.g.][]{Kamlesh20,Rajpurohit21a,Rajpurohit21}. 
 The halo in the Coma cluster offers  a unique chance to perform spatially resolved studies of the halo brightness and of its  connection between thermal and non-thermal plasma \citep[e.g][]{Govoni01,BrownRudnick11}. 
 So far, these studies have been inhibited by the lack of resolution, as 
 it is often necessary to convolve the radio images with large Gaussian beams to recover the full halo emission.\\
 LOFAR \citep{vanHaarlem}, thanks to its sensitivity and resolution, provides a step forward for precise measurements of the halo size and flux density, and for resolved studies. In this paper, we study the radio emission from the Coma cluster at 144\,MHz, using data from LoTSS (\citealt{Shimwell19}, Shimwell et al. in press) after ad-hoc reprocessing. 
 We also use published data, and reprocessed archival X-ray and radio observations, to perform a multi-wavelength study of the emission.\\
This paper is organized as follows: In Sec.~\ref{sec:data} we describe the observations and data reduction procedures, in Sec.~\ref{sec:diffuse} we present the main sources of diffuse emission, both known and newly discovered. The radio halo is analysed in Sec.~\ref{sec:halo_fit}, its spectrum in Sec. ~\ref{sec:haloSpectrum} and its correlation with the thermal gas are shown in Sec.~\ref{sec:radioXcorrelation} and \ref{sec:radioSZ}. In particular, in Sec. 5.3 and 6.4, the halo properties as discussed in the framework of turbulent re-acceleration models, and constraints to the model parameters are derived. In Sec.~\ref{sec:simulations}, we use cosmological MHD simulations to reproduce the thermal/non-thermal properties of the Coma cluster. In Sec. \ref{sec:haloFront}, the halo front is analysed and its origin is discussed in connection with the shock wave found by X-ray and SZ studies. We also analyse the emission detected between the relic and the radiogalaxy NGC4789 in Sec.\ref{sec:relicNAT}. Finally, results are discussed in 
sec.~\ref{sec:discussion}.
Throughout this paper, we use a $\Lambda$CDM cosmological model, with $\Omega_{\Lambda}=0.714$, $H_0=$69.6 km/s/Mpc. At the Coma redshift ($z=0.0231$) the angular to linear scale is 0.469 kpc/arcsec and the Luminosity distance $D_L=101.3$ Mpc.

\begin{figure*}
\centering
\includegraphics[width=2\columnwidth]{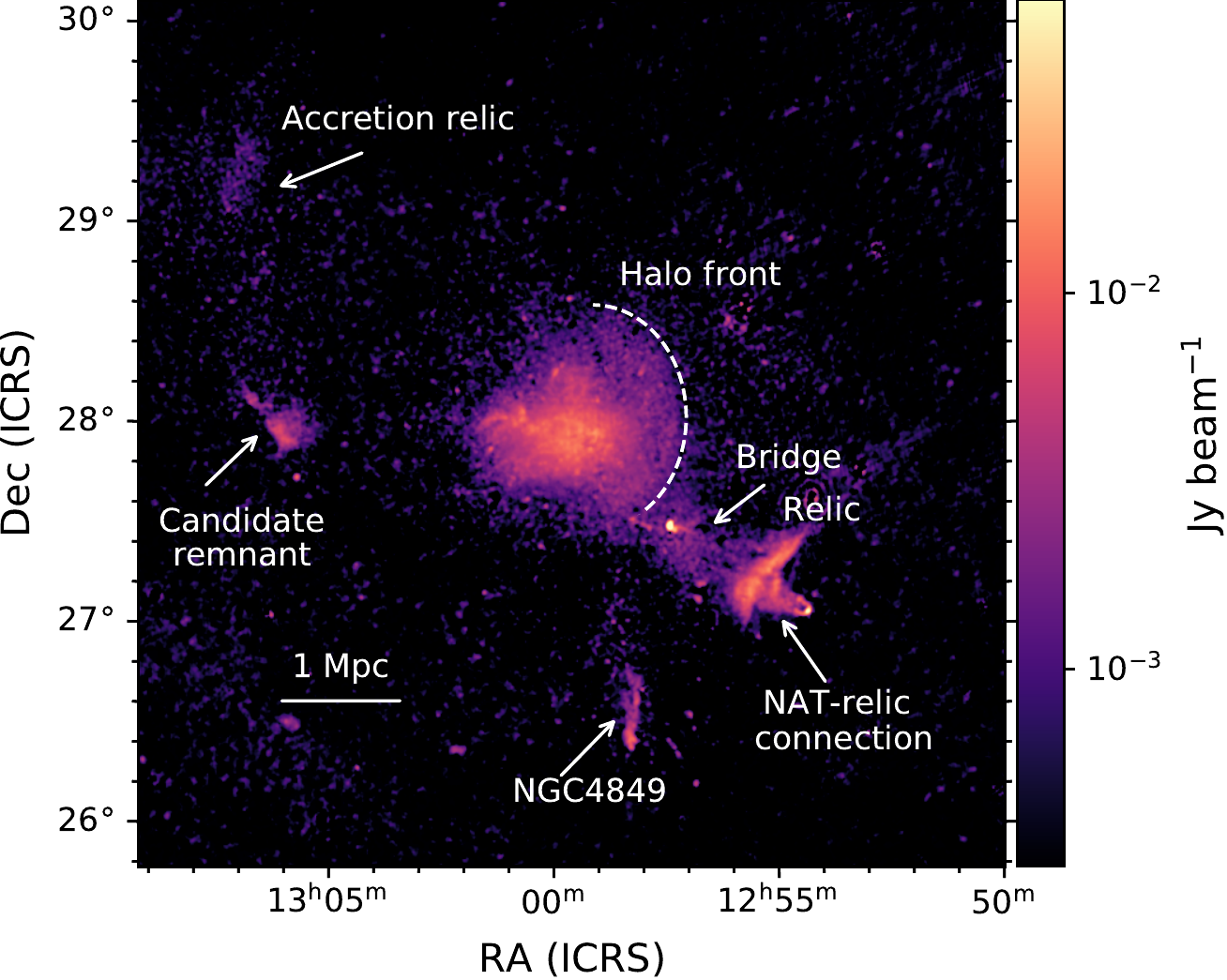}
\caption{LOFAR emission of the Coma field at the resolution of 1 arcmin. The candidate accretion relic, halo front, bridge, relic, and NAT-relic connection are labelled.}
\label{fig:Coma_fov}
\end{figure*}

\begin{figure*}
\centering
\includegraphics[width=2.1\columnwidth]{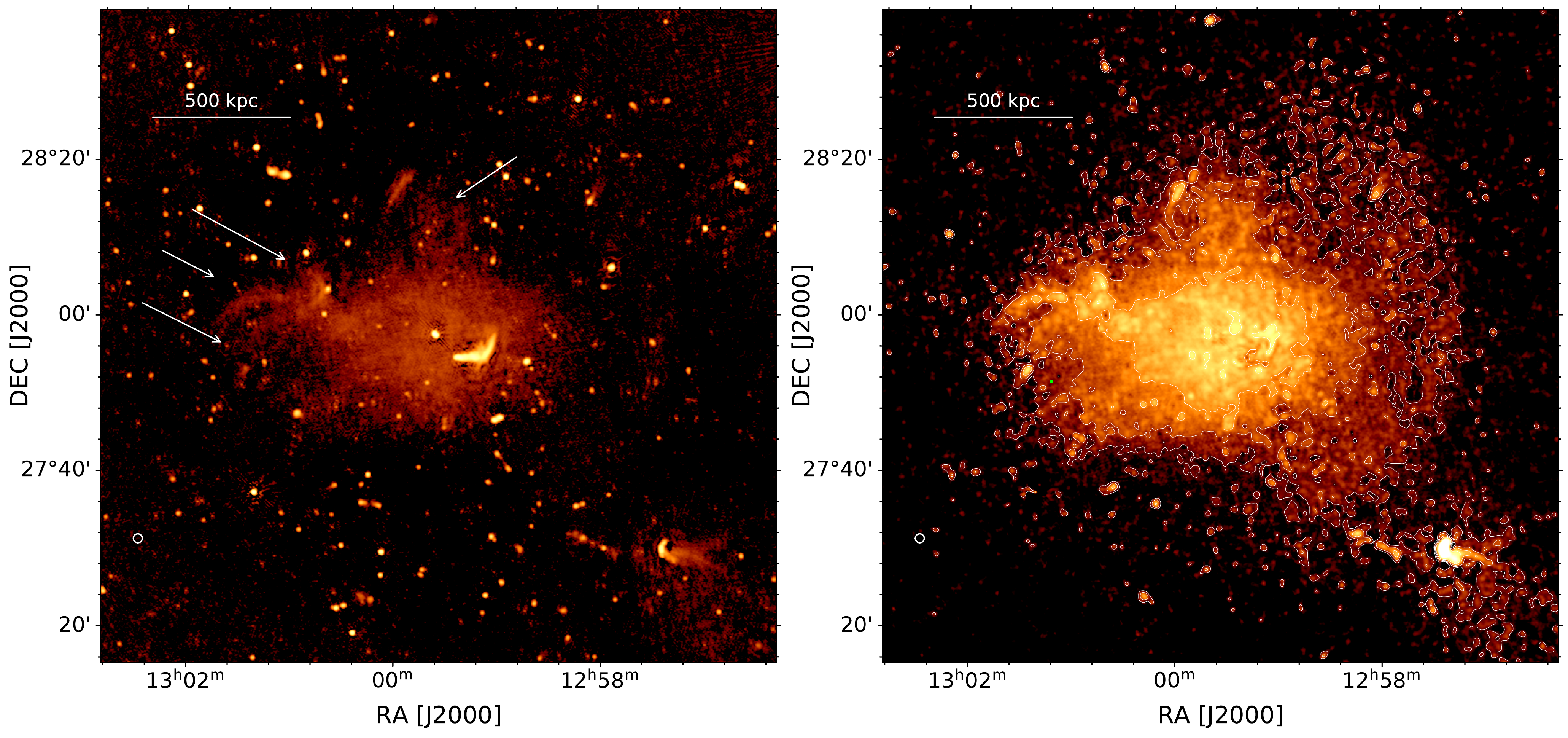}
\caption{Left panel: zoom into the radio halo from the 20\asec image, which is made imposing an inner UV cut (image LOFAR as LoTSS 20 in Tab. \ref{tab:images}). The halo core and the radio galaxies, both, of Coma and of the field are visible. The black bowl around the core indicates that large-scale diffuse emission is filtered out.
White arrows mark the filaments (3 arrows on the left) and the radio-loop (arrow on top-right). Right panel: same as left panel but from the 35\asec image, where all the baselines have been included in the image and the discrete sources have been subtracted (see text for details)
The outer halo and the halo front are well visible. Contours start at 3$\sigma_{\rm rms}$ and are spaced by a factor of 2.}
\label{fig:zoomHalo}
\end{figure*}

\begin{figure*}
\centering
\includegraphics[width=2.1\columnwidth]{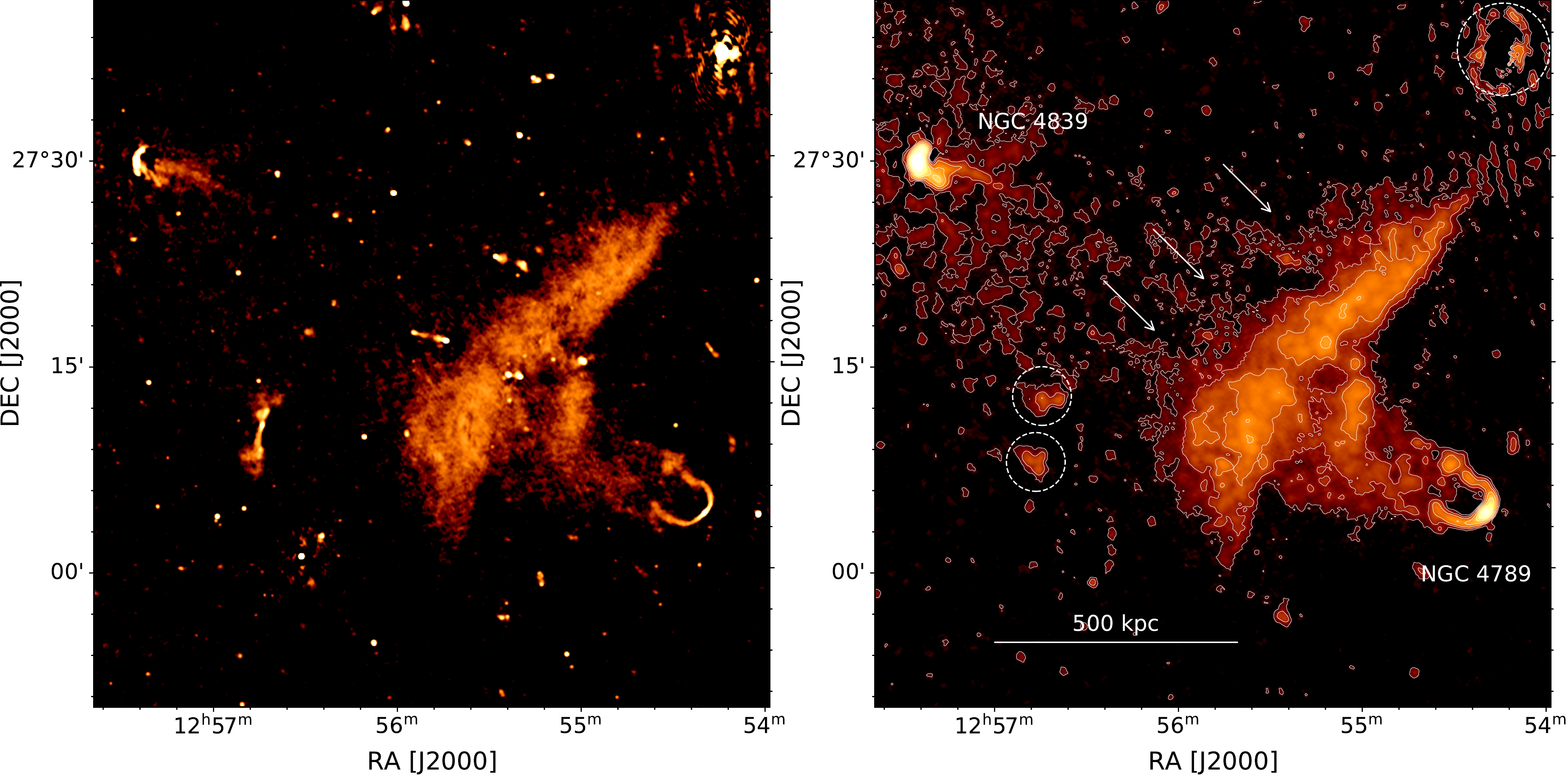}
\caption{Zoom into the radio relic region. Left panel:
  6\asec image showing the complex emission that links the source NGC 4789 to the relic emission. Right panel: 35\asec image. The source NGC 4789 is labelled. Arrows mark the position of the stripes that depart from the relic towards NGC4839. White dashed circles indicate residuals from the subtraction of the lobes of NGC4827, a Coma radiogalaxy, and Coma A. Contours start at 3 $\sigma_{\rm rms}$ and are spaced by a factor of 2.}
\label{fig:zoomRelic}
\end{figure*}

\section{Observations used in this work}
\label{sec:data} 

\subsection{LOFAR data  and data reduction}
The data used in this work are part of the LoTSS (\citealt{Shimwell19}, Shimwell et al. 2022) and consist of 2 pointings of 8~hr each taken with LOFAR (the LOw Frequency ARray, \citealt{vanHaarlem}) High Band Array antennas, in the DUAL\_INNER mode configuration. Each pointing is 8~hr long, book-ended by 10~min observations of a calibrator (3C196), used to correct for the ionospheric Faraday rotation, clock offsets, instrumental XX and YY phase offsets, and time-independent amplitude solutions. The pointings are specified in Tab. \ref{tab:obs} together with the distance from the central source NGC~4874, at the cluster centre.
Observations are centered at 144\,MHz and have a 48\,MHz total bandwidth. After pre-processing and direction-independent calibration, data are averaged into 24 visibility files, each having a bandwidth of 1.953\,MHz with a frequency resolution of 97.6\,kHz and a time resolution of 8s. For details about pre-processing and direction-independent calibration steps, we refer to \cite{Shimwell19}, where LoTSS data acquisition and processing are both explained in detail.
The direction-dependent calibration has been made using the LoTSS DR2 pipeline, but with a slightly different procedure to account for the large-scale emission present in the Coma field. Specifically, we have included all baselines in imaging and calibration, while the LoTSS DR2 pipeline applies an inner \emph{uv}-cut of the visibilities below 100~m to eliminate radio frequency interferences on the shortest baselines and filter out large-scale Galactic emission that would make the process of calibration and imaging more difficult. The two pointings have been calibrated separately and unrelated sources have been subtracted from the visibilities through a multi-step procedure, for which we refer the reader to \cite{Bonafede21}. Briefly, we first subtracted all the sources outside a radius of 1.5$^{\circ}$ centered on the cluster, using the model components obtained from an image at 20\asec resolution with no inner  \emph{uv}-cut, and using a threshold of 5$\sigma$ (0.75 mJy/beam). The subtracted data have been reimaged using an  \emph{uv}-cut of 300~m to filter out the diffuse emission from the ICM, and the model components have been subtracted using a threshold of  0.6 mJy/beam. Data have been reimaged again to check for the presence of residual emission from discrete sources using an \emph{uv}-cut of 100 m. Residual emission associated with Active Galacic Nuclei (AGN) and sources unrelated to the halo and relic emission has been identified and subtracted by the data. The sources NGC4789 and NGC4839 have not been subtracted in order to study how their emission is connected to the relic and bridge, respectively.
Two diffuse patches of emission to the West of Coma are still present and have been deconvolved in the imaging runs. They are discussed in the following sections.\\
The images used in this work have been done in a different way than in \cite{Bonafede21}, taking advantage of the new implementations that have been added to the imaging software in the last months. Imaging has been done with DDFacet \citep{Tasse18}, using the recently added features that allow to deconvolve large-scale emission in joint-deconvolution mode.
We used the Sub Space Deconvolution (SSD) algorithm (\citealt{Tasse18} and ref. therein) to better model the clean components. Typically, four major cycles were needed to achieve a noise-like residual map. 
The beam correction has been applied in the image plane, interpolating the beams of the two different pointings in the direction of Coma. 
Different images have been produced, as listed in Tab. \ref{tab:images}, at various resolution to highlight the emission from discrete sources and from the diffuse emission.
To align the flux scale to LoTSS, we have extracted the fluxes from the sources in the Coma field, and followed the same bootstrap procedure described in \citet{Hardcastle16}, which is based on the NVSS (NRAO VLA Sky Survey, \citealt{NVSS}).

\subsection{XMM-Newton and ROSAT data}
We used the \emph{XMM-Newton} Science Analysis System (SAS) v18.0.0 for data reduction. The ObsIDs we used are listed in Appendix \ref{appendix:obsid}. Event files from the MOS and pn detectors were generated from the observation data files with the tasks \texttt{emproc} and \texttt{epproc}. The out-of-time (OoT) events of pn were corrected following the user guide\footnote{\url{http://xmm-tools.cosmos.esa.int/external/xmm_user_support/documentation/sas_usg/USG/removingOoTimg.html}}. 
We used stacked Filter Wheel Closed (FWC) event files to generate non X-ray background (NXB) maps. For each ObsID, the FWC event files were reprojected using task \texttt{evproject} to match the observation. NXB maps were scaled to match the NXB level of each count image. For MOS, the scale factor was calculated using the ratio of out-of-field-of-view (OoFoV) count rates. For pn, even the OoFoV area can be contaminated by soft protons, and cannot be used for an accurate rescaling of the NXB \citep{Gastaldello17,Zhang20,Marelli21}. On the other hand, the pn instrumental background shows a similar long term variability as that of \emph{Chandra}'s ACIS-S3 chip; therefore we used this information for rescaling the pn NXB level in the observations (details are provided in Zhang et al. in prep).\\
The corresponding exposure maps were generated using the task \texttt{eexpmap} with parameter \texttt{withvignetting=yes}. Point sources were detected and filled by the tasks \texttt{wavdetect} and \texttt{dmfilth} in the $Chandra$ Interactive Analysis of Observations (CIAO) v4.13 package. We stacked individual count maps, exposure maps, and NXB maps, respectively, using the 0.5--2.0 keV energy band. We divided the NXB subtracted count image by the exposure map to generate the flux map. After removing the instrumental background and correcting for telescope vignetting, a constant sky background was further subtracted from the images. The level of this background was estimated from the median flux in an annulus spanning radii of 60-70 arcmin. Note that \citet{Mirakhor20} do detect a signal from the Coma ICM even at these large radii, albeit in a narrower energy interval of 0.7-1.2 keV where the signal to background is optimized. The sky background used here may therefore be slightly overestimated, but this does not have an impact on the radii of interest considered in our \emph{XMM-Newton} analysis (limited in this case mostly within the cluster's R$_{500}$). \\
We furthermore used the four archival observations of the Coma cluster performed in June 1991 by the ROSAT Position Sensitive Proportional Counter (PSPC). These pointings 
extend out to radii of 60-70 arcmin, totaling a clean exposure time of 78 ks. The data reduction was performed exactly as already described in \citet{Simionescu13,Bonafede21}. The low instrumental background of ROSAT makes these data an important complement to \emph{XMM-Newton} in the faint cluster outskirts.

\subsection{Planck data}
We used six Planck frequency maps acquired by the High Frequency Instrument (HFI) and the corresponding energy responses released in 2018 by the Planck collaboration \citep{Planck20}.  The average frequencies of the HFI maps are 100, 143, 217, 353, 545 and 857 GHz. The nominal angular resolution of the maps are 9.69, 7.30, 5.02, 4.94, 4.83, and 4.64 arcminutes, respectively. Following prescriptions of Galactic thermal dust studies performed by the Planck collaboration (Planck collaboration, 14), we corrected HFI maps for individual offsets that maximise their spatial correlation with neutral hydrogen density column measurements performed in regions of the sky characterised by their low dust emissivity and by the exclusion of prominent Sunyaev-Zel'dovich (SZ) sources. \\
This multi-frequency data set allowed us to map the thermal SZ Compton parameter in a four square degree sky area centred on Coma, following the spectral-imaging algorithm described in \citet{Baldi19}. Briefly, the Compton parameter, $y$, is jointly mapped in the wavelet space with the Cosmic Microwave Background anisotropies and with the Galactic thermal dust emissivity. This is achieved via a spatially-weighted likelihood approach that includes the smoothing effect of individual HFI beams in the reconstruction of B3-spline wavelet coefficients. In order to best restore anisotropic details, a curvelet transform is eventually computed from these wavelet coefficients, and denoised via a soft-thresholding that we parametrise as a function of local values of the noise standard deviation. The resulting $y$-map has a resolution (FWHM of the PSF) of 5 arcmin, i.e. a factor 2 higher than the one published in \cite{PlanckComa}. \\

\subsection{WSRT data}
We use data from the Westerbork Synthesis Radio Telescope (WSRT) at $\sim$ 325 MHz. Part of these data come from the observations published in \cite{Venturi90,Giovannini91}, that we have reimaged as explained in \cite{Bonafede21}.
In this work, we also use more recent WSRT observations published by \cite{BrownRudnick11} that recover a larger fraction of the radio halo, but are more affected by imaging and calibration artefacts due to the presence of the source Coma A, North of the relic. 
\cite{BrownRudnick11} observations are used here to study the spectral properties of the halo, while \cite{Venturi90,Giovannini91} observations are used here to study the spectral index in the relic region. In Tab. \ref{tab:images}, these images and their main properties are listed. We refer to the image published by \citet{BrownRudnick11} as WSRT H, as it is used to study the halo emission, and to the image published by \cite{Venturi90,Giovannini91} as WSRT R, as it is used to study the relic region.
 
\begin{table}
 \centering
\caption{ LOFAR observations details}
\begin{tabular}{l c c c c } \hline
 LoTSS  & RA & DEC & time  & dist of NGCC4874 \\
       pointing       &deg & deg & h     & deg \\
 P192$+$27   &   192.945 &	27.2272  &  8 &  1.88  \\
 P195$+$27   &   195.856  &	27.2426   &  8 & 1.11 \\
\hline
\end{tabular}
\label{tab:obs}
\end{table}

\begin{figure*}
\includegraphics[width=2.1\columnwidth]{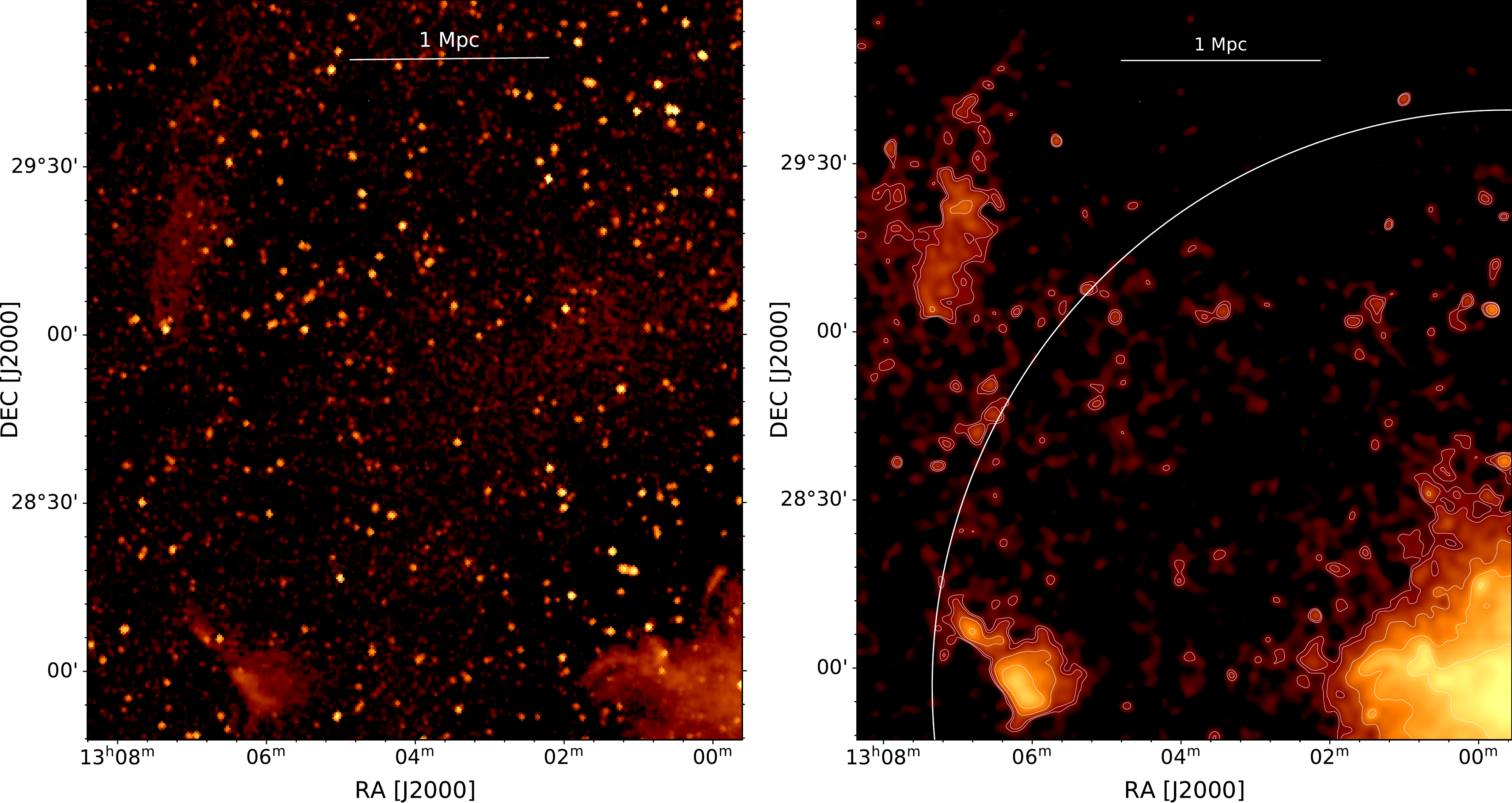}
\caption{Zoom into the region of the accretion relic. Left panel: 20\asec  image from LoTSS showing the diffuse emission of the accretion relic (top left), the candidate remnant source
(bottom left) and all the radio sources in the field. Right panel: same as left panel but from the 2$^{\prime}$ image. Contours are plotted at (3,4,8,16,32,64)$\sigma_{\rm{rms}}$. The white circle is centered on the Coma cluster centre and has a radius $r=R_{100}$. }
\label{fig:zoomAR}
\end{figure*}

\begin{figure*}
\includegraphics[width=2.1\columnwidth]{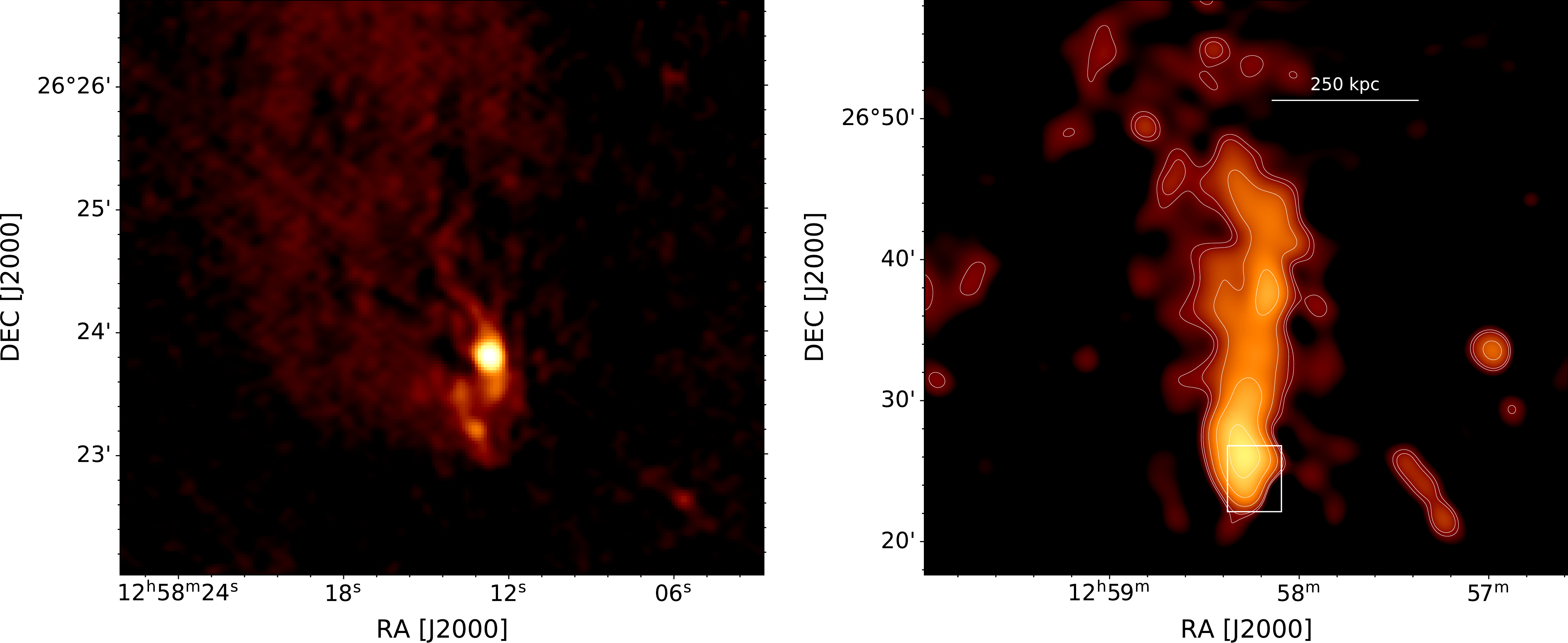}
\caption{Zoom into the region of NGC~4849, to the South of Coma. Left panel: 6\asec  image from LoTSS showing the core of NGC~4849. Right panel: 1$^{'}$ image showing the full extent of  the tail. The white box marks the region shown in the left panel. Contours start at 3 $\sigma$ and are spaced by a factor 2.
}
\label{fig:NGC4849}
\end{figure*}

\section{Components of the diffuse emission}
\label{sec:diffuse}

The radio emission from the Coma field consists of several different components, either associated with the ICM or originating from the interaction of the radio galaxies with the environment. 
In Fig.~\ref{fig:Coma_fov}, the diffuse emission from the Coma cluster field is shown, after the subtraction of radio galaxies and point sources. 
The most relevant features are labelled. As already known, the Coma cluster hosts a radio halo, a radio relic, and a bridge of low surface brightness emission connecting the two. In addition to these components, several new features are detected in our LOFAR observations. In this Section, we present the emission from the diffuse sources as imaged by LOFAR at 144\,MHz. An analysis of their properties that includes radio images at other frequencies and the comparison with the X-ray emission from the ICM is presented in the following sections for each source separately.

\subsection{The radio halo}

At 144\,MHz, the radio halo appears larger than at higher frequencies, with a LAS of 1.2 deg, measured East West, corresponding to $\sim$ 2\,Mpc (see Tab. \ref{tab:sources}). The halo appears to be composed of a central, bright core and a larger, weaker component that is asymmetrical and more pronounced towards the West (see Fig. \ref{fig:zoomHalo}). 
The inner portion of the halo, is what we define the ``halo core"\footnote{The halo core and outer halo are labelled in Fig. \ref{fig:corr_core_outer}}.
 This part of the halo is the one visible in the LoTSS images which impose an inner \emph{uv}-cut of $80 \lambda$, corresponding to $\sim 43'$. The 20\asec resolution of LoTSS, and here reproduced using the same \emph{uv}-range restriction, provides a detailed image of the inner portion of the halo. Clearly, the emission from the halo core is brighter than the rest, and its surface brightness is characterized by bright filaments of radio emission, marked with arrows in Fig.~\ref{fig:zoomHalo}. 
 Despite being the best studied radio halo, this is the first time that features such as these filaments are detected in its diffuse emission.\\
Outside the halo core, low surface brightness emission is detected, which is brighter towards the West. We call this emission the ``outer halo'' and discuss in Secs.~\ref{sec:halo_fit} and \ref{sec:radioXcorrelation} whether the halo core and the outer halo show different properties, as recently proposed for some cool-core clusters (\citealt{Savini_2018,Savini_2019,Biava21b,Riseley21}). 
The emission from the outer halo is only visible when baselines shorter than 80$\lambda$ are included in the image, and hence it is filtered out in the LoTSS images.\\

\begin{figure}
    \centering
    \includegraphics[width=\columnwidth]{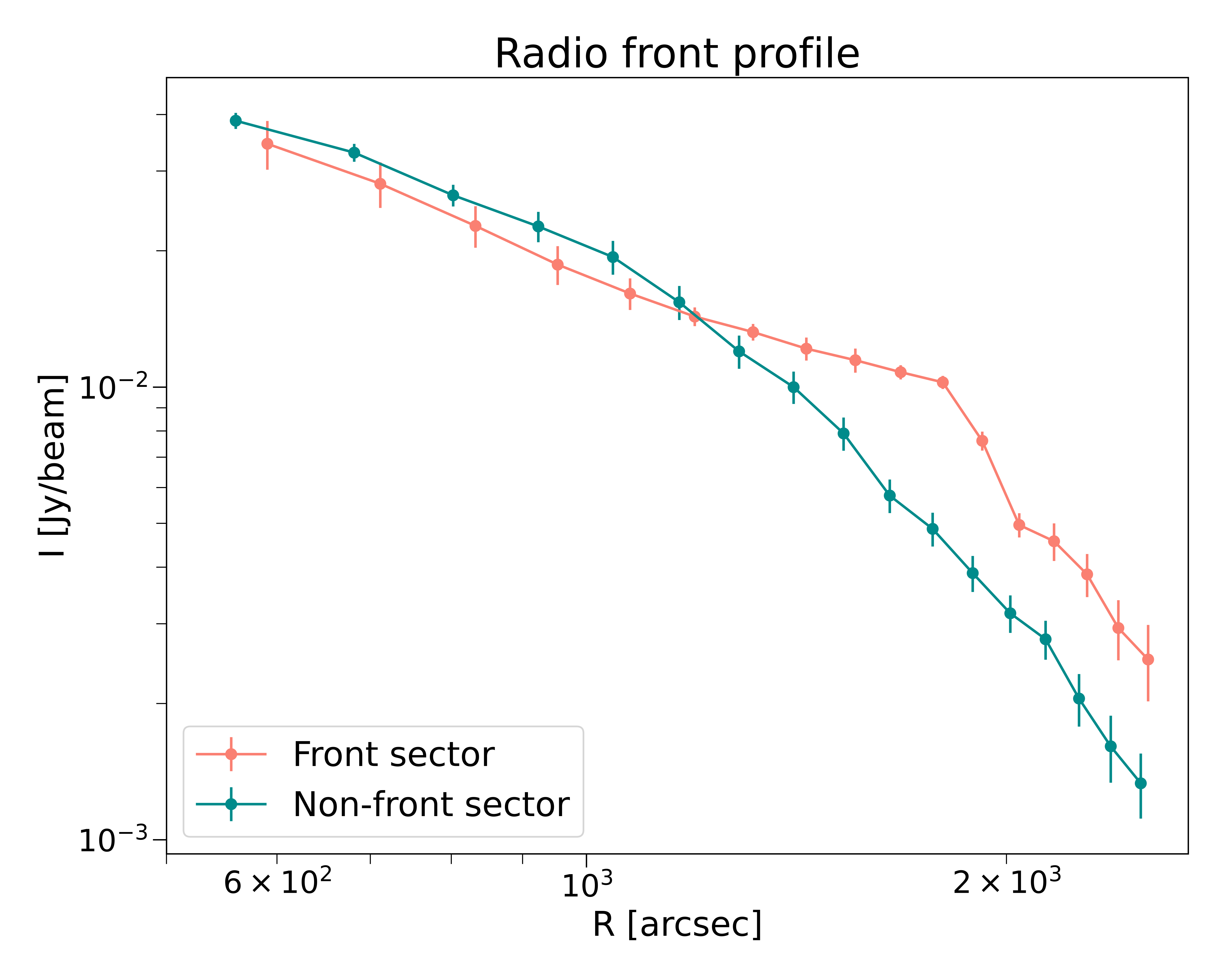}
    \caption{Radio profile of the Coma halo form the LOFAR image at 2$^{\prime}$ resolution across the halo front in comparison with the rest of the halo. Errorbars represent statistical errors only.   }
    \label{fig:profile_front}
\end{figure}

In the right panel of Fig.~\ref{fig:zoomHalo}, we show the halo in the LOFAR 35\asec  image, after the subtraction of the unrelated radio sources.
To the West, the  halo has a sharp edge coincident with the halo front (see Fig.~\ref{fig:Coma_fov} for labelling) already found by \citet{BrownRudnick11} and coincident with the shock front detected in the X-rays \citep{Simionescu13} and SZ \citep{PlanckComa}.
In Fig. \ref{fig:profile_front} we show the radio profile computed in annuli across the halo front in comparison with the rest of the cluster. The two sectors have  azimuth angles from 47$^{\circ}$ to 287$^{\circ}$ (Coma) and from 323$^{\circ}$ to 48$^{\circ}$ (Front, centered in RA=12:59:07, DEC=+28:01:31). The Southwest part of the halo, towards the radio bridge, has been excluded. While the average cluster profile shows a smooth radial decline, a sharp edge is visible around $\sim$1900\arcsec from the cluster centre, coincident with the radio front found by \cite{BrownRudnick11}.\\
Towards the southwest, the spherical front continues and its brightness becomes weaker and merges with the emission from the radio bridge. The total flux density of the radio halo, measured from the images above 2$\sigma_{\rm rms}$ is 10$\pm$2 Jy at 144\,MHz. The main properties of the radio halo are listed in Tab.~\ref{tab:sources}.

\subsection{The radio relic and the NAT-relic connection}
To the South-West of the Coma cluster, a radio relic has been discovered by \citet{Ballarati81} and studied by several authors afterwards \citep{Venturi90,Giovannini93,BrownRudnick11, OgreanComa, Bonafede13}. However, the presence of the bright source ComaA, at the north-west of the relic, has always made its study difficult. Residual calibration errors remain in the LOFAR image (see Fig.~\ref{fig:zoomRelic}) and make it difficult to determine whether the relic extension to the NW is real. We note that the same extension has also been detected by single dish observations at 1.4 GHz, where it also appears polarized \citep{BrownRudnick11}. Despite this, because of its questionable nature, we adopt a conservative approach and do not include it in our discussion of the relic.\\
The radio relic has a LAS of 38$'$, corresponding to $\sim$ 1.1\,Mpc at the Coma redshift. 
Its emission is connected on both sides to two head-tail radio galaxies, namely NGC~4839 to the North-East, and NGC~4789 to the South-West of the relic. From NGC4839, we detect diffuse emission that blends into the bridge and relic emission \citep{Bonafede21}. The surface brightness of the relic is not uniform, but composed of patchy and filamentary sub-structures (see Fig.~\ref{fig:zoomRelic}). Stripes of radio emission depart from the relic and are directed towards the bridge and NGC 4839. 
A similar stripe is detected also near the tail of NGC 4839, directed towards the relic. These features could be regions where the magnetic field has been amplified and/or where the plasma has been stripped from the tail of NGC~4789.
The surface brightness of these stripes is a factor 2-3 higher than the nearby emission, and their size is $\sim$ 11.7$'$ that corresponds to 300 kpc at the Coma's redshift.\\
Beyond the relic, towards the SW, the narrow angle tail galaxy (NAT) NGC 4789 appears connected to the relic, as already found by e.g. \citealt{Giovannini91}. The connection between the relic and NGC 4789 has been interpreted as the tail of NGC 4789 feeding the relic with radio plasma \citep[e.g.][]{EnsslinGK01}. In Sec.~\ref{sec:relicNAT}, we will investigate the possible origin of this emission. \\
\subsection{Diffuse emission from NGC4849}
To the South of Coma, we detect diffuse emission, elongated in the North-South direction, with a size of $\sim$27$^{\prime}$. This emission seems associated with the radiogalaxy NGC~4839, located at RA$\rm{=12h58m12.679s}$, DEC$\rm{=+26d23m48.77s}$, at redshift $z=0.01966$. Its angular size translates into a linear size of $\sim$650 kpc (see Fig. \ref{fig:NGC4849}). From the 6\asec resolution image, the core of NGC~4849 is visible, and a hint of jet emission in the North-South direction is also present. A second bright component is located at the South of the core, which could either be an unrelated radio source seen in projection, or the lobes of NGC~4849. The maximum extension of this tail in the WSRT image is $\sim$7$^{\prime}$. Using the WSRT H and ``LOFAR as WSRT H" images (see Sec. \ref{sec:data} and Tab. \ref{tab:images}) and taking into account the higher noise of the WSRT image in that region, we derive a 2-$\sigma$ limit for the spectral index of the tail $\alpha < -2$. Analysing the emission of this source is not the aim of this paper, we note that similar steep spectrum tails have been found in the outskirts of other clusters (e.g. Abell~1132 \citealt{Wilber17}, Abell~2255 \citealt{Botteon20}) likely tracing the motion of the galaxy in the ICM. The long tail of aged cosmic ray electrons (CRe) left behind by these sources during their motion provides seed electrons that could be re-accelerated by turbulence and shocks.

\subsection{Accretion relic}

After the subtraction of unrelated sources, two extended patches of diffuse emission are visible at the periphery of Coma, to its North-East and East (see Fig.~\ref{fig:zoomAR}). The diffuse component at the Eastern side of Coma is not associated to any Coma cluster galaxy \citep{PizzoTesi}. \citet{PizzoTesi}
also noticed that this source lies at the crossroad of two filaments of galaxies, pointing towards the clusters A~2197 and A~2199, respectively, and suggest that this emission could be due to the accretion of matter towards the Coma cluster. Our images have a higher resolution and allow us to recognize a double-lobe structure that resembles a radiogalaxy. However, no core is obviously associated with it in the 6\asec and 20\asec images (see Fig.~\ref{fig:zoomAR}).  We tentatively label this source as a ``candidate remnant", but we note that its classification remains uncertain. Indeed, given its large angular size ($\sim$ 27$^{'}$) and the lack of an optical identification at the Coma cluster redshift, it would have a very large linear size if it was in the background of Coma. \\
The other diffuse patch in the NE part of Coma is entirely new  and has an arc-like morphology, and a weak uniform brightness. This emission does not appear connected to any discrete source (see Fig.~\ref{fig:zoomAR}). 
When convolving the image to a resolution of 2 arcmin, it has a largest angular size of $\sim$ 55', that would correspond to 1.5\,Mpc at the cluster redshift, and it is located at $\sim \, 2.1^{\circ}$ from the cluster centre.
At the Coma redshift, this distance corresponds to $\sim$3.6 Mpc, while the cluster virial radius\footnote{$R_{100}$ is considered here to be a good approximation of the virial radius, as the cluster redshift is 0.023} is $\sim 2.9$ Mpc.
Since recent studies \citep[e.g.,][]{Malavasi20} have found an intergalactic filament of galaxies in the NW direction from the Coma cluster, this source could be connected to the cluster and its large-scale environment. Hence, we consider it plausible that this is an ``accretion relic", though more data supporting our hypothesis are needed.
Considering that accretion shocks on these scales should be characterized by a strong Mach number ($\mathcal{M} \gg 5$, e.g. \citealt{Hong14}), a strong prediction in this case would be that the radio spectrum of this emission should be $\alpha \propto \nu^{-1}$, and also characterized by a large degree of polarisation. 
The properties of this accretion relic are listed in Table \ref{tab:sources}.

\begin{table*}
\centering
\caption{Images used in this work}
\begin{tabular}{l  c c c c}\hline
Image name &  Freq & Resolution & $\sigma_{\rm rms}$ & Fig. \\
     & MHz &  & mJy/beam  &            \\
LOFAR as LoTSS 6    & 144      &  6 \asec $\times$ 6 \asec  & 0.1  & Fig. \ref{fig:zoomRelic} \\
LOFAR as LoTSS 20    & 144      &  20 \asec $\times$ 20 \asec  & 0.15  & Fig. \ref{fig:zoomHalo} \\

LOFAR 35\asec       & 144 &  35 \asec $\times$ 35 \asec &  0.2  & Fig. \ref{fig:zoomHalo} (top panel),  Fig. \ref{fig:zoomRelic}. \\
 LOFAR 1$'$     & 144 & 60  \asec $\times$ 60 \asec & 0.4 & Fig. \ref{fig:Coma_fov}\\
&&&& \\
WSRT H  &  342 &  134\asec $\times$ 68 \asec   & 0.4 & \citealt{BrownRudnick11} \\
LOFAR as WSRT H  & 144  & 134\asec $\times$ 68 \asec  & 1.5  & Fig. \ref{fig:spix_front} \\
WSRT R   &  326 &   150\asec $\times$ 100 \asec &  1.2  & \citealt{Giovannini91,Bonafede21}\\
LOFAR as WSRT R & 144 & 150\asec $\times$ 100 \asec & 1 &  \\
\hline
\end{tabular}
\label{tab:images}
\end{table*}

\begin{table*}
\centering
\caption{Source properties at 144 MHz}
\begin{tabular}{l c c c c} 
\hline
 Source name & dist & LAS & $S_{\rm{144 \, MHz}}$  & $P_{\rm{144 \, MHz}}$ \\
             &     &     &    Jy &    W/Hz \\
Halo    &      -     &   71$'$ - 2.00 Mpc  &   12$\pm$2  & 1.5 $\pm 0.2 \times 10^{25}$ \\
Accretion relic   &   2.1$^{\circ}$ - 3.55 Mpc   &  55$'$ - 1.5 Mpc    &  0.47 $\pm$ 0.07   &  5.7$\pm 0.9 \times 10^{23}$\\
Relic                & 73$'$ - 2.0 Mpc  &  38$'$ - 1.1 Mpc  &  2.4 $\pm$ 0.4   & 3.0$\pm 0.4 \times 10^{24}$ \\
Relic-NAT connection &  79$'$ - 2.3 Mpc& 10$'$ - 280 kpc   &  0.7$\pm$0.1      &  9$\pm 1 \times 10^{23}$    \\
\hline
\multicolumn{5}{l}{Col. 1: Name of the diffuse source; Col 2: Distance from the cluster centre; Col 3: Largest angular size measured  }\\
\multicolumn{5}{l}{above the 2$\sigma_{mrs}$ contour Col. 4: FLux density measured above the 2$\sigma_{\rm rms}$ contour; Col 5: Radio power at 144 MHz}\\
\multicolumn{5}{l}{The K-correction is applied assuming $\alpha=-1, -1, -1.2, -1.4$}\\
\multicolumn{5}{l}{for the halo, accretion relic, relic, relic-NAT connection, respectively.}\\
\end{tabular}
\label{tab:sources}
\end{table*}

\section{The radio halo profile}
\label{sec:halo_fit}
To characterize the halo properties, we have fitted its surface brightness profile adopting the approach firstly proposed by \citet{Murgia09} and more recently generalized by \citet{Boxelaar_2021} to account for asymmetric halo shapes.
The main novelty of this procedure is that the profiles are fitted to a two-dimensional image directly, using MCMC to explore the parameter space, rather than to a radially averaged profile. In addition, the fitting procedure by \citet{Boxelaar_2021} allows one to fit also elliptical and skewed (asymmetric) models. 
We refer to  \citet{Boxelaar_2021} for a detailed explanation and summarize here the relevant parameters.
The surface brightness model is given by:
\begin{equation}
    I(r) = I_0 \exp^{-G(r)} ,
    \label{eq:exp}
\end{equation}
where $I_0$ is the central surface brightness and $G(r)$ a radial function. $G(r)= \left( \frac{|r^2|}{r_e^2}\right)^{0.5}$ for the circular model, while $G(r)= \left( \frac{x^2}{r_1^2} + \frac{y^2}{r_2^2}\right)^{0.5}$ for the elliptical models, where $r_e$ is the characteristic e-folding radius, and $r^2=x^2 + y^2$. The skewed model allows for an off-center maximum of the brightness distribution, and is characterized by four different scale radii ($r_1$, $r_2$, $r_3$, $r_4$) and by an angle to describe the asymmetric brightness distribution.\\
We have investigated these three profiles (circular, elliptical, and skewed), and results of the fit are given in Table \ref{tab:fit_halo}.
In this table, we only list the statistical error, while the systematic one is 15\% of the listed flux density, and due to the uncertainty on the absolute flux scale.
All models give a consistent total radio power ($P_{144 \rm{MHz}} \sim 1.47 \times 10^{25} \, \rm{W \, Hz^{-1}}$ computed within 3 times the e-folding radius) and central radio brightness ($I_0 \sim 5  \, \mu \rm{Jy/arcsec^{2}}$). We note that the flux density and consequently the power measured by the fit are in perfect agreement with the estimate derived from the images above 2$\sigma_{\rm rms}$ (see Sec. \ref{sec:diffuse}).
As the reduced-$\chi^2$ values are similar in the skewed and elliptical model, we consider the latter in the following analysis.\\

\begin{figure*}
    \centering
    \includegraphics[width=\columnwidth]{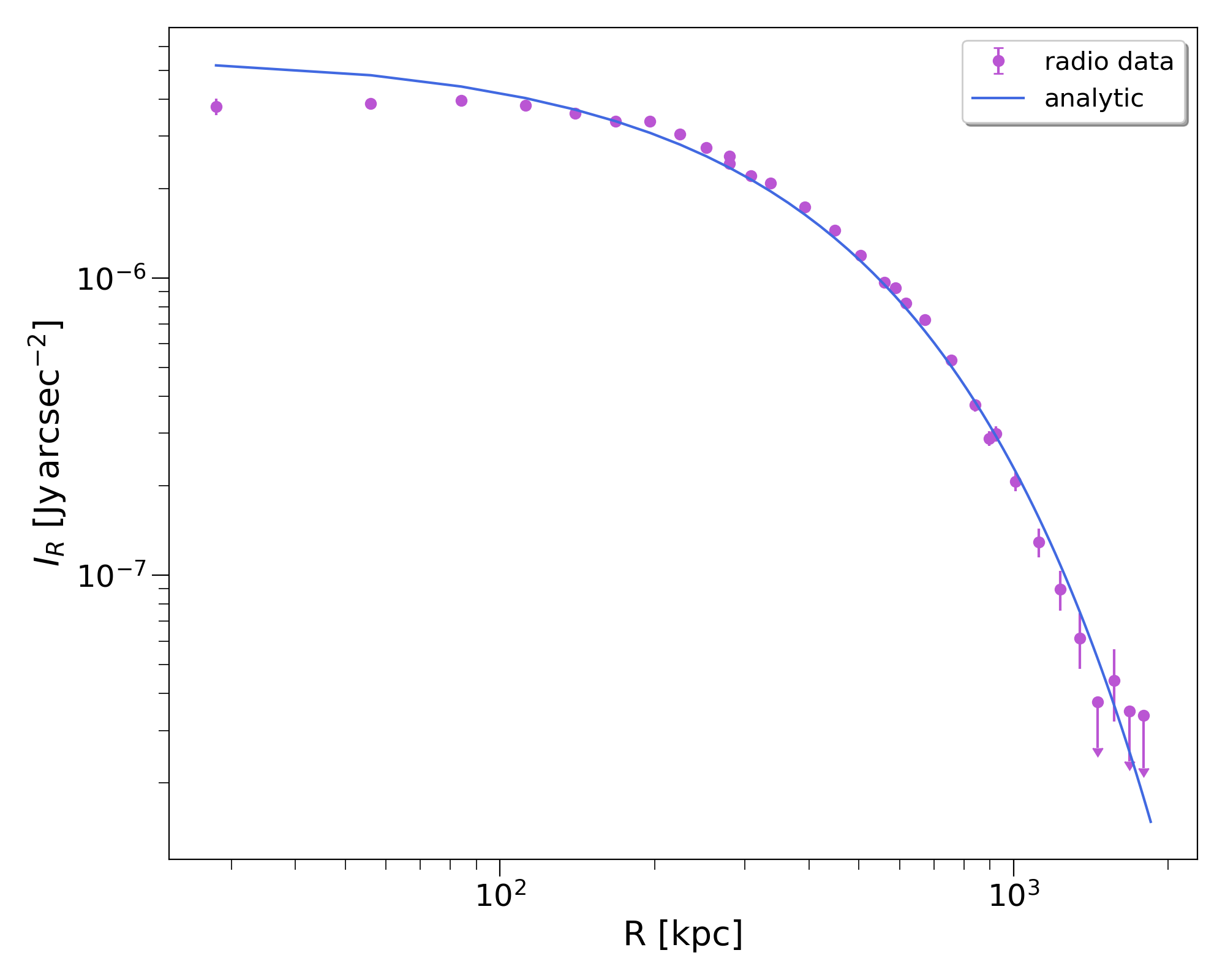}
    \includegraphics[width=\columnwidth]{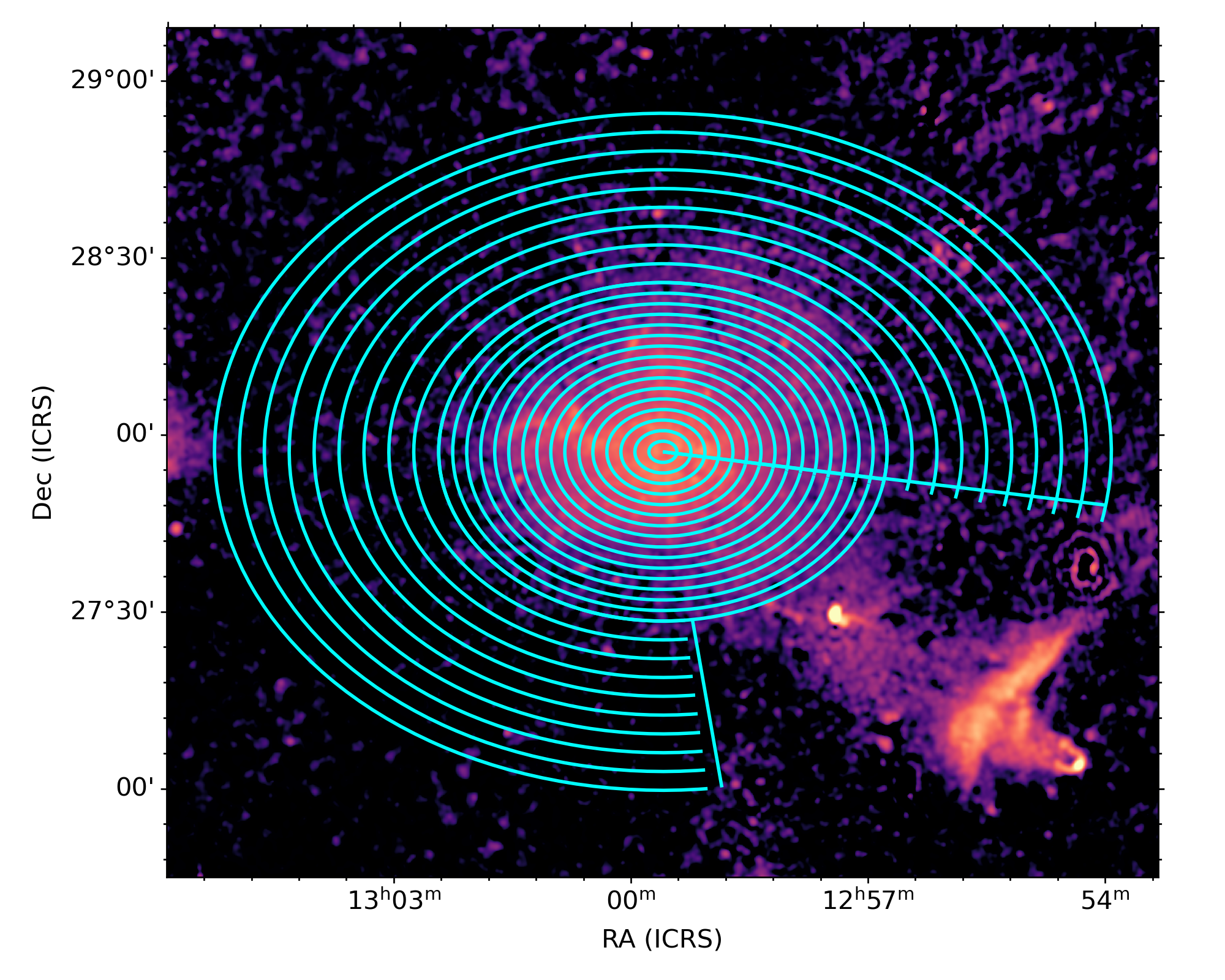}
    \caption{Left panel: bullets are the radio average brightness profile of the radio halo, computed within elliptical annuli having major and minor axis sub-multiples and multiples of $r_1$ and $r_2$, i.e. the minor and major axis of the elliptical exponential fit (see Tab. \ref{tab:fit_halo}). The SW sector of the halo has been blanked for ellipses with major and minor axis larger than 3$r_1$ and 3$r_2$, respectively, to exclude the bridge region.
    Errorbars represent the error on the mean, upper limits at 3$\sigma $ are plotted as arrows. The width of the annuli goes from 1$^{\prime}$ in the center to 4$^{\prime}$ in the outer annuli, to improve the sensitivity to low surface brightness emission in the outer parts of the halo.
    The blue line refers to the best fit elliptical model, see Tab. \ref{tab:fit_halo}.
    Right panel: radio image of the Coma cluster at 1$^{\prime}$ resolution. The inner cyan ellipse has major and minor axis equal to 3$r_{1}$ and 3$r_{2}$ respectively, and elliptical annuli are spaced by 1$^{\prime}$. The outer elliptical bins have a width of 4$^{\prime}$ and and trace ellipses out to 6$r_{1}$ and 6$r_{2}$, showing the SW region that has been excluded from the analysis because of the bridge. }
    \label{fig:halo_profile}
\end{figure*}

In Fig. \ref{fig:halo_profile}, we show the radial profile of the halo brightness at 144 MHz. We have computed the mean of the radio brightness and its error within elliptical annuli having a width than changes progressively from 1$^{'}$ in the centre  to 4$^{'}$ in the outer regions, to maximise the resolution at the centre and the sensitivity to low surface-brightness emission in the halo peripheral regions. The bridge region s excluded starting from elliptical annuli with major and minor axis larger than 3$r_1$ and 3$r_2$, respectively.
We considered upper limits the values where we have a mean smaller than 3 times the rms noise. From this plot, we derive that we have detected the halo emission up to $r\sim 1.3$ Mpc, that corresponds to $\sim$ 4.5$r_1$ and $r_2$ (see Tab. \ref{tab:fit_halo}).\\
In the following, we refer to the halo core as the emission contained within an ellipse having as major and minor axis $r_1$ and $r_2$, respectively, and to the 
outer halo as the emission contained within  4$r_1$ and 4$r_2$, excluding the halo core. We note that the halo core and outer halo are well represented by a single exponential model and do not need to be considered as two separate components, if we consider the average surface brightness of the halo emission. However, as we show in the following analysis, they are characterized by different properties.\\
We applied the same fitting procedure to the WSRT image (WSRT H in Tab. \ref{tab:images}), after blanking the sources that were contaminating the emission. We only attempted an elliptical and circular model fit. 
Results are listed in the bottom part of Table \ref{tab:fit_halo}. Both fits give the same total flux density and radio power, and have the same $\chi^2_r$ value, smaller than 1, possibly indicating that the errors are being overestimated.
We note that the $r_1$ and $r_2$ values are slightly smaller than those found in the LOFAR image, indicating a more peaked profile of the radio emission, hence of the emitting CRe at higher frequencies.

\begin{table*}
\caption{Radio halo 2D fit}
\begin{tabular}{lccccccccc}
\hline
\hline
\multicolumn{10}{l}{LOFAR - 144 MHz}\\
\hline
Halo model & $\chi_r^2$ & $I_0$ & $r_1$  & $r_2$& $r_3$& $r_4$ & angle &  S$_ {144, \rm{MHz}}$ & $P_{144, \rm{MHz}}$\\
    &       &$\mu \rm{Jy}/\rm{arcsec}^2$ & kpc&kpc&kpc&kpc& deg & Jy & $\rm{ 10^{25} W \, Hz^{-1}}$ \\
Circular    & 1.7   &     5.42 $\pm$ 0.04 &   310 $\pm$ 1 &&&&& 12.20 $\pm$ 0.04 & 1.470 $\pm$ 0.002 \\ 
Elliptical & 1.6     & 5.50$\pm$ 0.02 &  355 $\pm$ 1 & 268 $\pm$ 1 & & &  & 12.21 $\pm$ 0.04 &1.470 $\pm$ 0.005 \\
Skewed & 1.6 & 5.46 $\pm$ 0.01 & 337$\pm$2 & 368$\pm$2 & 207$\pm$2 & 342$\pm$ 2 & 2.99$\pm$ 0.001 & 12.35 $\pm$ 0.05 & 1.490$\pm$ 0.05 \\
\hline\hline
\multicolumn{10}{l}{WSRT - 342 MHz}\\
\hline
Halo model & $\chi_r^2$ & $I_0$ & $r_1$  & $r_2$& $r_3$& $r_4$ & angle &  S$_ {342, \rm{MHz}}$ & $P_{342, \rm{MHz}}$\\
    &       &$\mu \rm{Jy}/\rm{arcsec}^2$ & kpc&kpc&kpc&kpc& deg & Jy & $\rm{ 10^{25} W \, Hz^{-1}}$ \\

Circular  &0.7  & 3.52$\pm$0.08 & 255$\pm$4 &&&&& 5.3 $\pm$ 0.1 & 0.64 $\pm$ 0.01 \\

Elliptical & 0.7 & 3.54$\pm$0.08 & 268$\pm$5 & 240$\pm$6 & & & & 5.3$\pm$ 0.1 & 0.64 $\pm$ 0.01 \\
\hline\hline
\multicolumn{10}{l}{Col 1: Model used; Col 2: Reduced $\chi^2$ value; Col 3: Central brightness of the fit; Col 4 -7: e-folding radii; }\\
\multicolumn{10}{l}{Col 8: Angle for the skewed halo model fit; Col 9: Total halo power at 144 MHz computed within 3$r_e$.}\\
\multicolumn{10}{l}{Only statistical fit errors are shown in the table}
\end{tabular}
\label{tab:fit_halo}
\end{table*}

\begin{figure*}
    \centering
    \includegraphics[width=\columnwidth]{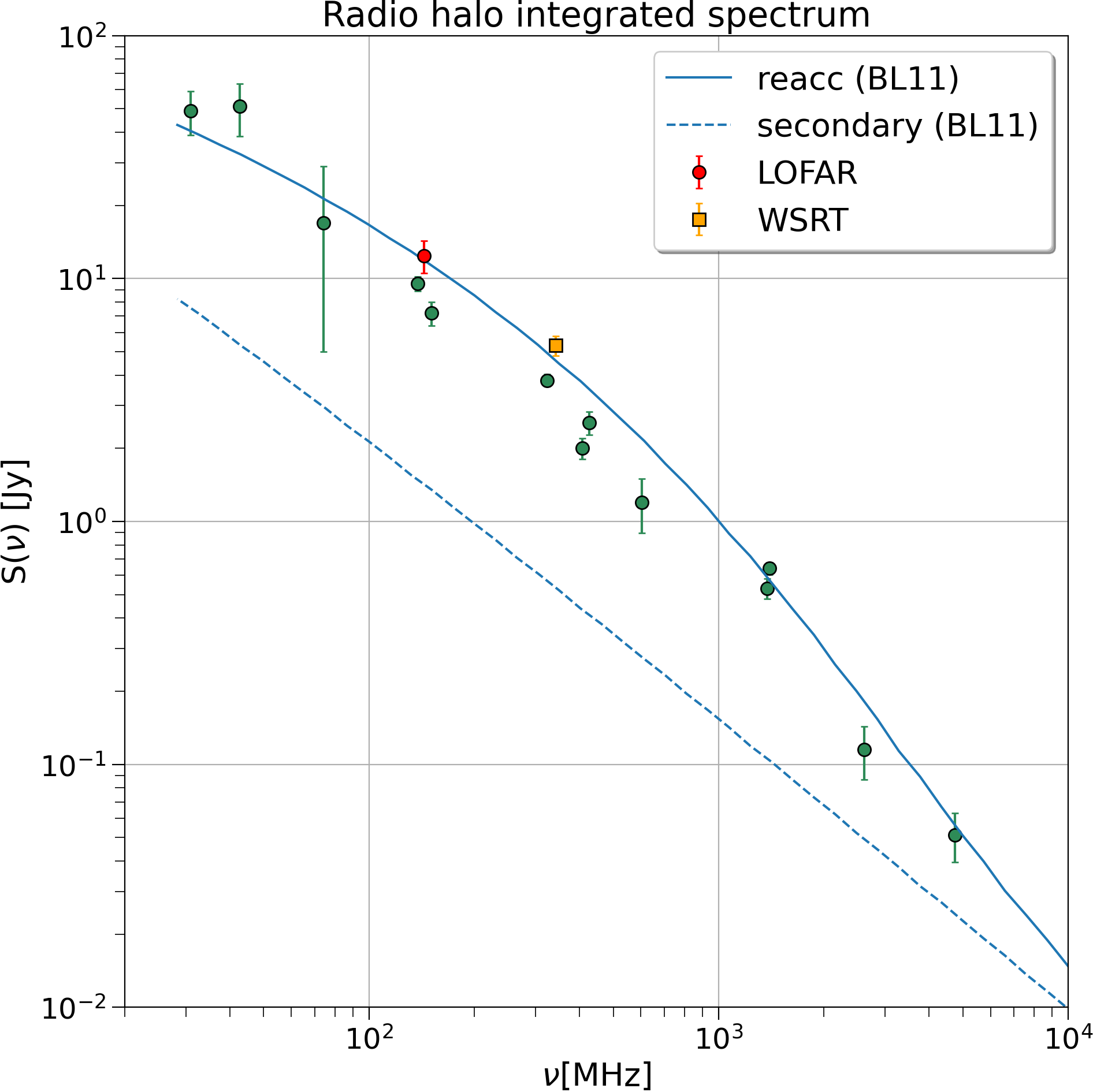}
    \includegraphics[width=1.01\columnwidth]{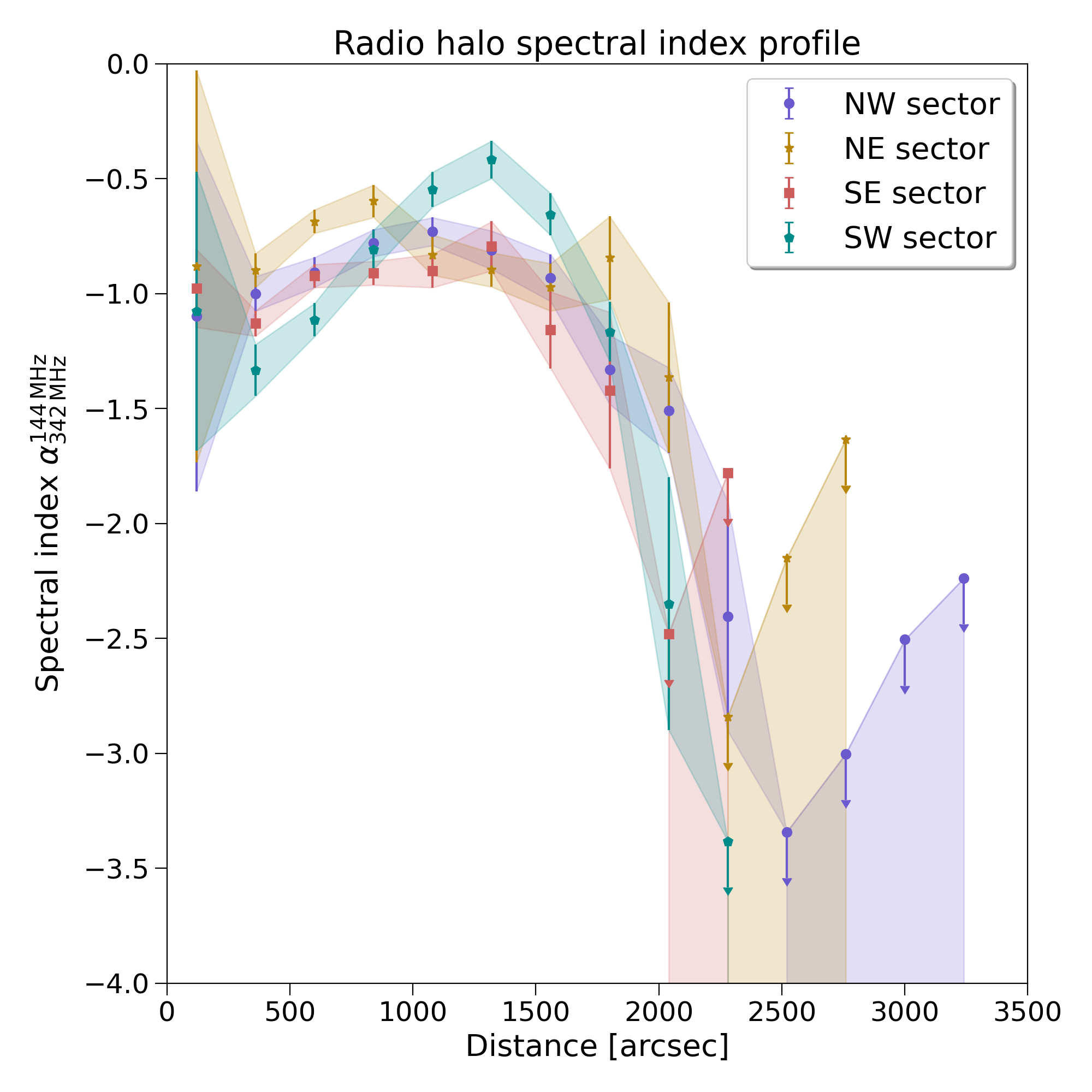}
\caption{Left: Integrated spectrum of the radio halo from literature data (green points) from  \cite{Thierbach03} and ref. therein, as corrected and re-scaled by \cite{Brunetti13}. The new LOFAR measurement (red point) and the WSRT measurement (orange box) are shown, measured within 3 times the effective radii as derived by the halo fit (see text for details). The continuous line shows  expectations from a model where secondary particles are reaccelerated by compressive turbulence \citep{BL11}. The dashed line shows the spectrum computed from secondary emission, after 200 Myrs since turbulent re-acceleration has been switched off \citep{BL11}. Right: spectral index radial profile computed in the different sectors, as listed in the legend. Arrows are 3$\sigma$ upper limits, only statistical errors are shown, the values of the spectral index are also affected by the flux calibration uncertainties of WSRT and LOFAR (10\% and 15\%, respectively) that would contribute with an additional error of 0.2.}
\label{fig:haloSpectrum}
\end{figure*}

\section{Spectral properties of the radio halo}
\label{sec:haloSpectrum}
\subsection{The integrated spectrum}
Using literature results, and the halo flux density at 144 and 342\,MHz, we can constrain the low-frequency part of the radio halo spectrum. The spectrum of radio halos provides important information about the underlying re-acceleration mechanism. Indeed, it allowed in the past to conclude that a very steep spectrum ($\alpha < -1.5$, also called ultra-steep spectrum halos) cannot be produced by hadronic models \citep[e.g.,][]{Brunetti08}.\\
Using the flux densities that result from the fits, we can measure the spectral index between 144\,MHz and 342\,MHz yielding $\alpha = - 1.0 \pm 0.2$. Though residuals from sources could still be present, the fitting procedure we have used should minimise that contribution \citep{Boxelaar_2021}, hence, a considerable impact to the whole halo emission is unlikely. We find that the spectrum is slightly flatter than previously reported in the literature, although still consistent within the errors.
The spectrum of the halo in the Coma cluster has been extensively studied in the literature \citep[e.g.][]{Giovannini93, Thierbach03}, using both interferometric images and single-dish data. At low frequencies, the contribution of radio galaxies is difficult to account for, for mainly two reasons: (i) the low resolution of observations published so far, (ii) the larger extent of tailed radio galaxies whose emission blends with the halo emission. Our sensitive and high-resolution LOFAR images allow us to alleviate both of these problems. In addition, we now have a more accurate method to estimate the halo flux density. 
We have collected the data available in the literature, i.e., those presented in \citet{Thierbach03}, and re-scaled to the same absolute flux scale as in \citet{Brunetti13},
and added our measurement at 144\,MHz. In Fig.~\ref{fig:haloSpectrum} we show the spectrum of the halo.
Green bullets refer to values taken from \citet{Brunetti13}, and do not refer to the same aperture radius. 
The red and orange point refer to  the LOFAR and WSRT map computed within the same aperture (corresponding to 3 e-folding radii of the LOFAR image, see Tab. \ref{tab:fit_halo}).
In Fig. \ref{fig:haloSpectrum}, left panel, we show one possible spectrum that would be produced by re-acceleration models.
However, for a proper derivation of the halo spectral properties, one should compare the flux densities from images done with the same \emph{uv}-range, the same procedure for compact source subtraction, and using the same aperture radius. We also note that using a fitting algorithm, as the one proposed by \cite{Boxelaar_2021} and used here, would also minimise the effect of different noise obtained at different frequencies.\\
Recently, \cite{Rajpurohit2022} have shown that the hint for a spectral break claimed in the radio relic of Abell 2256 below 1.4\,GHz \citep{Trasatti15} is not confirmed once the analysis is performed with  matching \emph{uv}-coverage and  unrelated sources  are subtracted properly. In the case of the Coma halo, we note that the high-frequency measurements are almost a factor 10 below the value extrapolated from a power-law at low frequencies. However, the exact shape of the spectrum could be affected by the effects mentioned above. In addition, it is important to get a precise flux density at frequencies below 100\,MHz to characterize the integrated radio spectrum. Here, data from the LOFAR Low-Band Antennas (LBA) will provide powerful constraints.
\subsection{Spectral index profile}
Using WSRT and LOFAR HBA data, we can obtain a radial profile of the halo spectral index. A steepening of the radial profile has been found by \citet{Giovannini93} using data between 326 MHz and 1.38 GHz. They found that the radio halo has a smooth spectrum in the central regions (inner 480 \asec radius) with $\alpha \sim -0.8$, and a steeper value ($\alpha \sim -1.2$ down to $-1.8$ in the outer regions).\\
Using WSRT and LOFAR HBA images, we can now compute the spectral index profile out to larger distances.
We have re-imaged the LOFAR data using only baselines larger than 40m, i.e. the shortest WSRT baseline, and convolved the LOFAR image to the same resolution as the WSRT image. The sources embedded in the diffuse emission are subtracted from the WSRT image, however, as some residuals were still present, we applied the multi-filtering technique described in \citet{Rudnick2002} and blanked the WSRT image wherever the filtered image was above 3 mJy/beam. We blanked the LOFAR image accordingly.
We have divided the radio halo into 4 sectors (NW, NE, SE, and SW) and computed the mean spectral index $\alpha$ in elliptical annuli having the major and minor axis proportional to $r_1$ and $r_2$, up to the maximum distance where the halo is detected (see Sec. \ref{sec:halo_fit} and Fig. \ref{fig:halo_profile}). \\
The 4 outermost annuli of the SW sector have been removed to exclude the bridge region. We have computed the spectral index in each annulus, and considered upper limits the annuli that have a total flux density smaller than 3$\sigma_{\rm rms} \times \sqrt(N_{\rm{Beams}}) $, with $N_{\rm{beams}}$ being the number of independent beams sampled in each annulus. As the radio halo is more extended in the LOFAR image than in the WSRT image, we mainly derive upper limits at distances larger than $\sim$ 2000\asec (950 kpc).
The radial trend of the spectral index is shown in Fig. \ref{fig:haloSpectrum}, right panel. We note that  values in some annuli are surprisingly flat, possibly because of residual contamination from unrelated sources. The increasing values of the upper limit in the outer annulli are due to the larger area sampled by the annuli.
All the sectors show a spectral index that becomes steeper at distances of $\sim$1500\asec from the cluster centre. 
In the SW sector, we also clearly detect a steepening towards the cluster centre. A similar, though less pronounced, trend is also observed in the other sectors, though we note that in the NE sector the radial profile of $\alpha$ is more complex, and no clear trend at distances smaller than $r \sim 2000$\asec can be established. The SW sector is affected by the passage of the NGC~4839 group, hence is it possible that different physical conditions are present there. Overall, we can cocnlude that the spectral trend is  characterized by clear steepening at the cluster outskirts, and a mild steepening towards the cluster centre. \\
In the next Sec., we will discuss the physical implications of these results in the framework of turbulent re-acceleration models.

\subsection{Radial variations of the spectral index and re-acceleration models}
In the presence of a break in the integrated spectrum of radio halos, homogeneous re-acceleration models predict that increasingly steeper spectra will be seen at increasing distance from the cluster center \citep{Brunetti01}. 
Assuming homogeneous conditions, the frequency at which steepening occurs, $\nu_s$, is proportional to:
\begin{equation}
\nu_s \propto \tau_{acc}^{-2} \frac{B}{ (B^2+ B_{IC}^2)^2}
\end{equation}
$\tau_{acc}$ is the re-acceleration time (see \citealt{CassanoBrunetti05}, \citealt{BrunettiLazarian07}) that depends on the assumed turbulent properties and re-acceleration mechanism. $B_{IC}$ is the inverse Compton equivalent magnetic field.
In the case of a constant  $\tau_{acc}$, the steepening frequency depends only on the magnetic field strength.  
To better follow the discussion below, let us define a critical magnetic field value
$$
B_{cr}=\frac{B_{IC}}{ \sqrt{3}} \sim 2 \mu G.
$$
As long as $B<B_{cr}$, one expects to see a radial steepening of the spectral index at distances larger and larger from the cluster centre as we move towards lower observing frequencies. This is what we observe beyond a radius of $\sim 30 ^{\prime}$.  \citet{Giovannini93} also detected a steepening between the higher frequency pair 325~MHz and 1.38~GHz, beyond a radius of  $\sim 8^{\prime}$ from the cluster centre. While we do expect more dramatic steepening at the higher frequencies, whether these two sets of measurements are consistent with a single physical model requires further investigation.\\
Moreover, we detect for the first time a steepening of the spectral index towards the cluster centre. This can also be explained in the framework of homogeneous re-acceleration models, if in the cluster centre we have $B>B_{cr}$.  Assuming a magnetic field profile as derived from RM studies ( $B\propto B_0 n_e^{0.5)}$ , \citealt{Bonafede10}), we have indeed a central magnetic field of $B_0 \sim 5 \mu$G, hence  $B>B_{cr}$.
We note that a similar steepening towards the cluster centre should be visible also in the higher frequency spectral index map by \citet{Giovannini93}. However, in that work the authors only report a spectral index trend through a line passing from the cluster from SE to NW. It is possible that a radial analysis similar to the one we present here would show the same trend. Alternatively, one should think of ad-hoc re-acceleration conditions that make this steepening visible only at low frequency. Future observations at higher ferquencies could shed light on this point.\\
We can conclude that the data presented in this work, together with literature data by \citet{Giovannini93} and \citet{Bonafede10} provide a coherent picture with the expectations from homogeneous turbulent re-acceleration models \citep{Brunetti01}, though the spectral index trend in the cluster centre leaves some open questions that could be addressed by future observations.\\
\subsubsection{Towards a constrain of the re-acceleration model parameters}
\label{sec:haloSpectrum_theo}
Constraining the model parameters, such as $\tau_{acc}$, would require a detailed 3D modeling and a precise constrain on the radial position of $\nu_s$. However, we can try to make first order calculations to see whether the qualitative coherent picture outlined above is quantitatively supported.
\citet{Giovannini93} have detected a spectral steepening at $\sim$8$^{\prime}$ between 326 MHz and 1.38 GHz. 
Assuming a central magnetic field $B_0=4.7 \mu$G, one would expect to detect the steepening between 144 and 342 MHz where the magnetic field is  $B\sim 0.7 \mu$G, i.e. at $r \sim 2$ Mpc ($\sim 71^{\prime}$) from the cluster centre. Instead, the steepening is detected at $r\sim$ 30$^{\prime}$ from the cluster centre, i.e. a factor 2 closer to the expected location
This may suggest that $\tau_{acc}$ is not constant and increasing with the distance from the cluster centre.
Though this result is not surprising from a theoretical point of view, it would be the first time that data support this claim.
Instead of a constant $\tau_{acc}$, we can make a step further and assume a constant turbulent Mach number $\mathcal{M}_t$. In this case, the steepening frequency is proportional to:
\begin{equation}
  \nu_s \propto T^{2a} \frac{B}{ (B^2+ B_{ICM}^2)^2}  
\end{equation}
where T is the cluster temperature and $a$ is a constant that is $a=1$ for 
re-acceleration via Transit-Time-Damping mechanism with compressive turbulence, and $a=1.5$ for non-resonant second-order acceleration with solenoidal turbulence\footnote{in this case we consider also a constant Alfv\'en velocity, based on the scaling $B^2 \propto n$.} \citep{BrunettiLazarian16}. 
In order to explain the spectral index steepening at $r\sim 30^{\prime}$ from the cluster centre, the temperature should be $\sim$35-45\% lower at a distance of 30$^{\prime'}$  than at a distance of  8$^{\prime}$, where the steepening is detected at higher frequency.
We note that this temperature drop is consistent with the temperature profile found by \citet{Simionescu13}\footnote{we take as a reference the profile extracted along the E sector of their analysis, which is the only one not contaminated by the W shock and by the merger with the NGC4839 group.}.
We stress again that the calculations we have just performed do not allow us to make any claim, as long as $\nu_s$ is not precisely determined from data and projection effects are not taken into account. However, they show that the qualitative picture outlined above is not in at odds with first order quantitative estimates.\\
\begin{figure}
    \centering
    \includegraphics[width=\columnwidth]{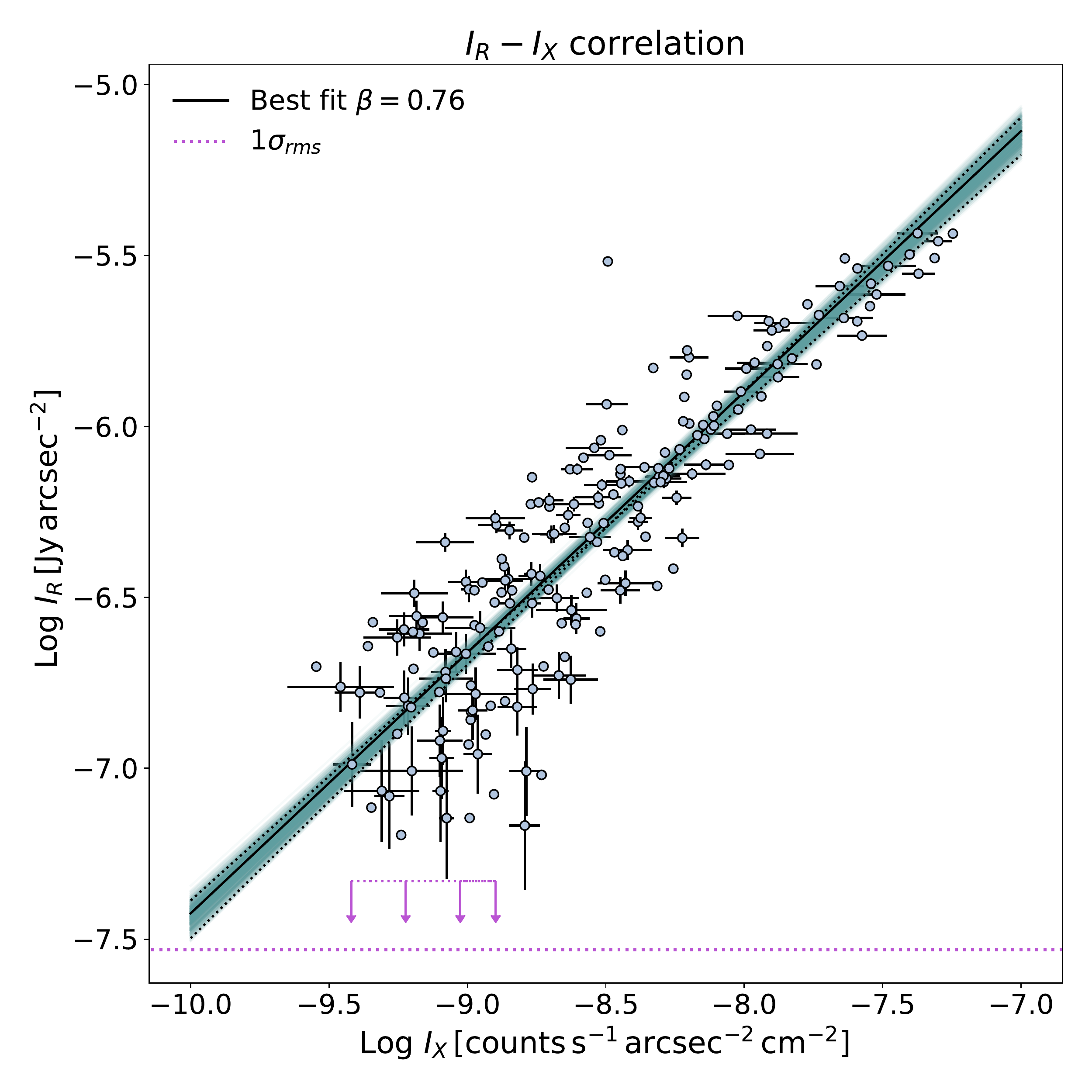}
    \caption{$I_R - I_X$ correlation computed from the 6$'$ image using cells equally spaced by 6$'$. Arrows mark the 2$\sigma_{\rm rms}$ upper limits, the black continuous line shows the best-fit line, dotted lines show the 10\% and 90\% slopes for the posterior distribution of $\beta$. The magenta dotted horizontal line marks the 1$\sigma_{\rm rms}$. All data are computed within 2400\asec from the cluster centre. Here and in the following correlation plots, errorbars are plotted every second point for clarity reason.}
    \label{fig:corr_trend}
\end{figure}

 \begin{table}
\caption{Radio - X-ray correlation for different Gaussian smoothing lengths.}
\begin{tabular}{c c c c}
\hline\hline
Gaussian Beam  & $\beta$ & 10\% -- 90\% & $\rho_{P}$  \\
FWHM   & && \\
1$^{\prime}$      &0.64       &  0.65 --  0.63   &   0.86 \\
2 $^{\prime}$        &  0.65  &  0.66 --  0.63   &   0.88 \\
3$^{\prime}$         & 0.68   &  0.70 --  0.66     &  0.89 \\
4 $^{\prime}$         & 0.70  &  0.73 --  0.68       &   0.88 \\
5 $^{\prime}$         &  0.74 &  0.78 --   0.70   &    0.89 \\
6 $^{\prime}$         &  0.76 &  0.81 --   0.72   &    0.88 \\
\hline\hline
\multicolumn{4}{l}{Col 1: FWHM of the smoothing Gaussian or restoring beam; }\\
\multicolumn{4}{l}{Col 2: Best-fit slope; Col 3: 10th and 90th percentile of the }\\
\multicolumn{4}{l}{posterior distribution for $\beta$. Col 4: Pearson correlation}\\
\multicolumn{4}{l}{coefficient}\\

\end{tabular}
\label{tab:radioX}
\end{table}

\section{Thermal and non-thermal correlations in the radio halo }
\label{sec:radioXcorrelation}
\subsection{Point-to-point analysis}
Investigating the point-to-point correlation between the radio and the X-ray surface brightness can give important information about the relation between the thermal and non-thermal components of the ICM. Also, it has the potential of constraining the mechanism responsible for the radio emission. \citet{Govoni01} have first investigated this correlation for a small sample of radio halos, finding a sub-linear scaling of the radio brightness with respect to the X-ray brightness. More recently, \citet{Botteon20, Rajpurohit21, Rajpurohit21a, Ignesti20} have investigated the same correlation for radio halos and mini halos, respectively, finding that halos tend to have a sub-linear or linear scaling, and mini halos show linear or super-linear behaviours.\\
Since the Coma radio halo  is the one for which most spatially resolved and multi-wavelength data are available, it is important to establish the statistical relation between its thermal and non-thermal components.\\
We have investigated the thermal to non-thermal correlation for the Coma cluster, fitting the radio ($I_R$) and X-ray ($I_X$) surface brightness in log-log space, according to:
\begin{equation}
\log I_R= \beta \log I_X + \gamma
\end{equation}
where  $\beta$ is the correlation slope.  
We have used a hierarchical Bayesian model \citep{Kelly07}, that allows us to perform linear regression of $I_R$ on $I_X$ accounting for 
intrinsic scatter in the regression relationship, possibly correlated  measurement errors, and selection effects (e.g., Malmquist bias). Using this method, we have derived a likelihood function for the data. We consider the mean of the posterior distribution as the best-fit slope. Following \cite{Botteon20,Bonafede21}, we considered as upper limits the radio values that are below 2$\sigma_{\rm rms}$.\\
Despite the fact that we detect the radio halo up to a distance of $\sim$1.3 Mpc, the analysis of the radio-X ray correlation is limited by the extent of the \emph{XMM-Newton} mosaic. In particular, because of soft-proton contamination, the analysis is restricted to a distance of 2400\asec ($\sim$ 1.1 Mpc) from the cluster centre.
To gain sensitivity towards the low surface brightness emission of the halo outskirts, we have convolved the radio image with Gaussian beams having FWHM of 1,2,3,4,5, and 6 arcminutes. We have computed the mean of the radio and X-ray brightness in square boxes having an area equal to a Gaussian beam of the radio image, and computed the fit using X-ray images smoothed at the same resolution. We have considered upper limits the values that are below 2$\sigma_{\rm rms}$, i.e. twice the noise of the radio image. 
The results of the fits are listed in Tab. \ref{tab:radioX}.
The image at 6$'$ resolution allows to recover the outermost regions of the halo keeping the highest possible resolution. 
$I_X$ and $I_R$ are positively correlated with a slope $\beta=0.76^{+0.05}_{-0.04}$ and a Pearson correlation coefficient $\rho_p=0.89$. 
 We note that the finest grid (boxes spaced by 1$^{\prime}$) recovers a slope similar to the one initially found by \citet{Govoni01}. The slope increases as we gain sensitivity to the low surface brightness emission that characterizes the outermost regions of the radio halo. This is suggesting that the slope of the correlation is not constant throughout the radio halo, with the slope $\beta$ increasing when low surface brightness emission is added. \\

\subsection{Correlations in the halo core and outer halo}
\begin{figure*}
\centering
\includegraphics[width=1.5\columnwidth]{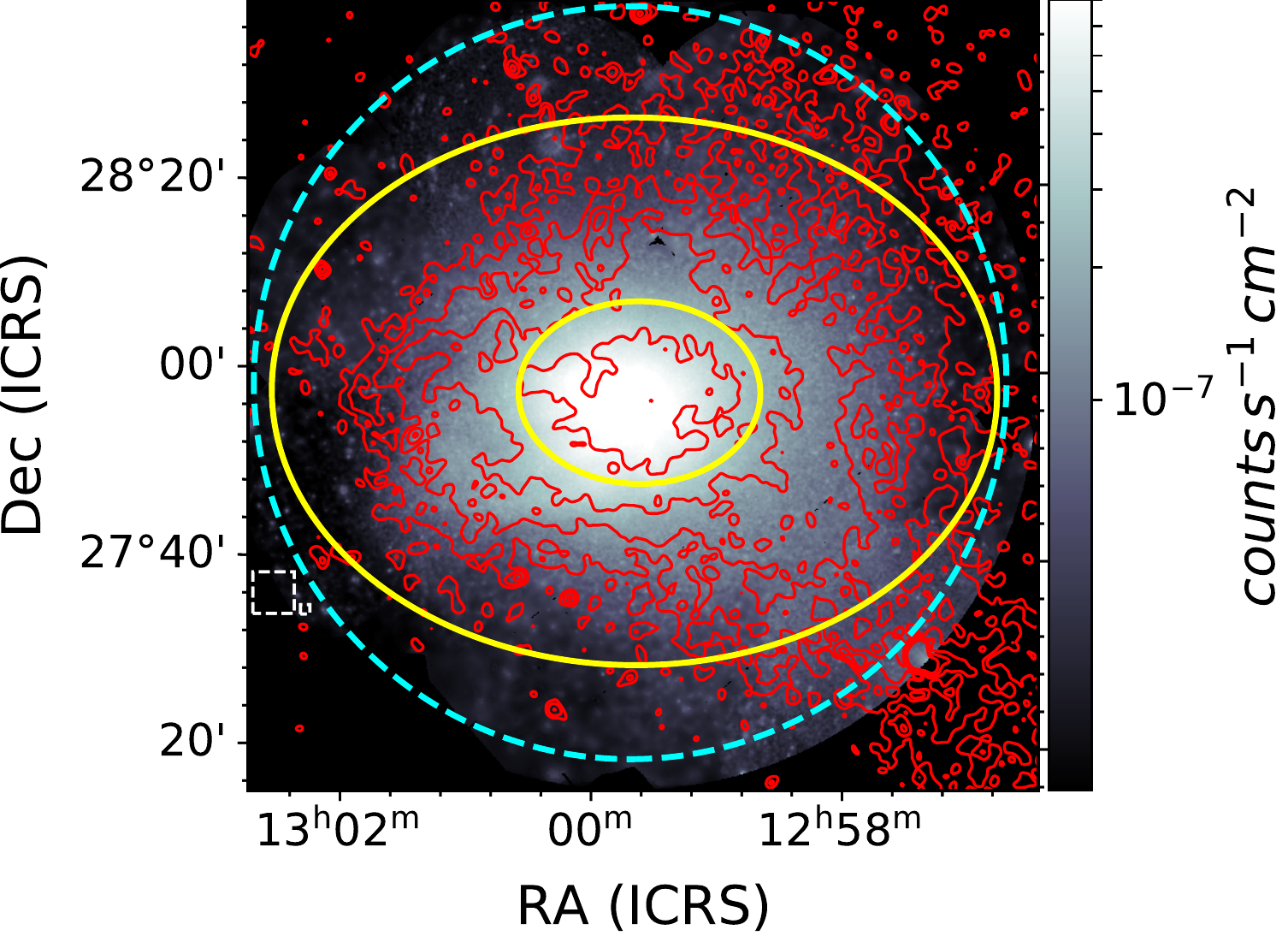}
\caption{Left panel: Colors are the X-ray emission from the Coma field from XMM-Newton observations. Contours show the radio emission at 1$^{'}$ resolution, starting at 3$\sigma_{\rm rms}$ and increasing by a factor 2.  The major axes of the ellipses are $r_1$ and $r_2$ (inner ellipse) and $3 r_1$ and 3 $r_2$ (outer ellipse).  $r_1$ and $r_2$ are listed in Tab. \ref{tab:fit_halo}. The inner ellipse is the region considered for the ``halo core". The cyan dotted circle mark the region that is not contaminated by soft X-ray protons.}
\label{fig:corr_core_outer}
\end{figure*}

Our  analysis suggests that the point-to-point $I_R-I_X$ correlation may be different in the halo central regions and in its outskirts.\\
Recent low-frequency observations have found radio emission in galaxy clusters that can be interpreted as the coexistence of a mini halo in the cluster core and a giant halo on larger scales (\citealt{Savini_2019}, \citealt{Biava21b}). 
A different $I_R - I_X$ trend is found in the core and in the outer part of the halo of these clusters, with a super-linear scaling in the mini halo region and a sub-linear scaling in the outer part (\citealt{Biava21b}, Lusetti et al., in prep).\\
Although those clusters have a cool-core and a radio halo with a steep spectrum, we note that a change in the $I_R$ - $I_X$ correlation slope for radio halos has never been explicitly investigated in the literature. 
In the cluster Abell 2142, a radio halo with two components has been found \citep{Venturi17}, yet no analysis of the radio and X-ray correlation has been performed so far. Also, in the cluster Abell 2744, a multi-component halo has been discovered with a different radio-X ray correlation slope for the northern and southern components \citep{Kamlesh2744}.\\
Since a point-to-point analysis of radio and X-ray surface brightness has been so far presented only for a few clusters, we investigate whether or not the different trends observed in the core and in the outer part of the Coma halo can be a common property of radio halos. 
We have repeated the analysis described in Sec.~\ref{sec:radioXcorrelation} considering the ``halo core" and the ``outer halo" separately (see Fig.~\ref{fig:corr_core_outer}). 
For the halo core, we have used the finest grid of 1$^{\prime}$ cell size, while for the outer halo we have used the 6$^{\prime}$ grid, to recover the faintest emission.\\
We find that the slope of the ``halo core'' is $\beta=0.41^{+0.04}$, while for the ``outer halo" 
we find $\beta=0.76 \pm {+0.05}$.\\
Hence, we can conclude that in the ``halo core" we find a flatter slope for the $I_R - I_X$ correlation, with respect to the outer part,  which is the opposite trend found in cool-core clusters that host a mini halo and a halo-type component. (e.g. RXC~J1720.1$+$2638, \citealt{Biava21b}; Abell 1413, Lusetti et al, in prep). We note that this is the first time that a change in the $I_R-I_X$ correlation has been investigated, hence it could be a common property of radio halos.
This result indicates that the inner part of the Coma radio halo has different properties than mini halos observed in more relaxed clusters,  likely indicating different local plasma conditions. In particular, the detection of a sub-linear trend in the brightest part of the halo hints at a negligible contribution  to the halo central emission from the hadronic mechanism.

\begin{figure}
    \centering
    \includegraphics[width=\columnwidth]{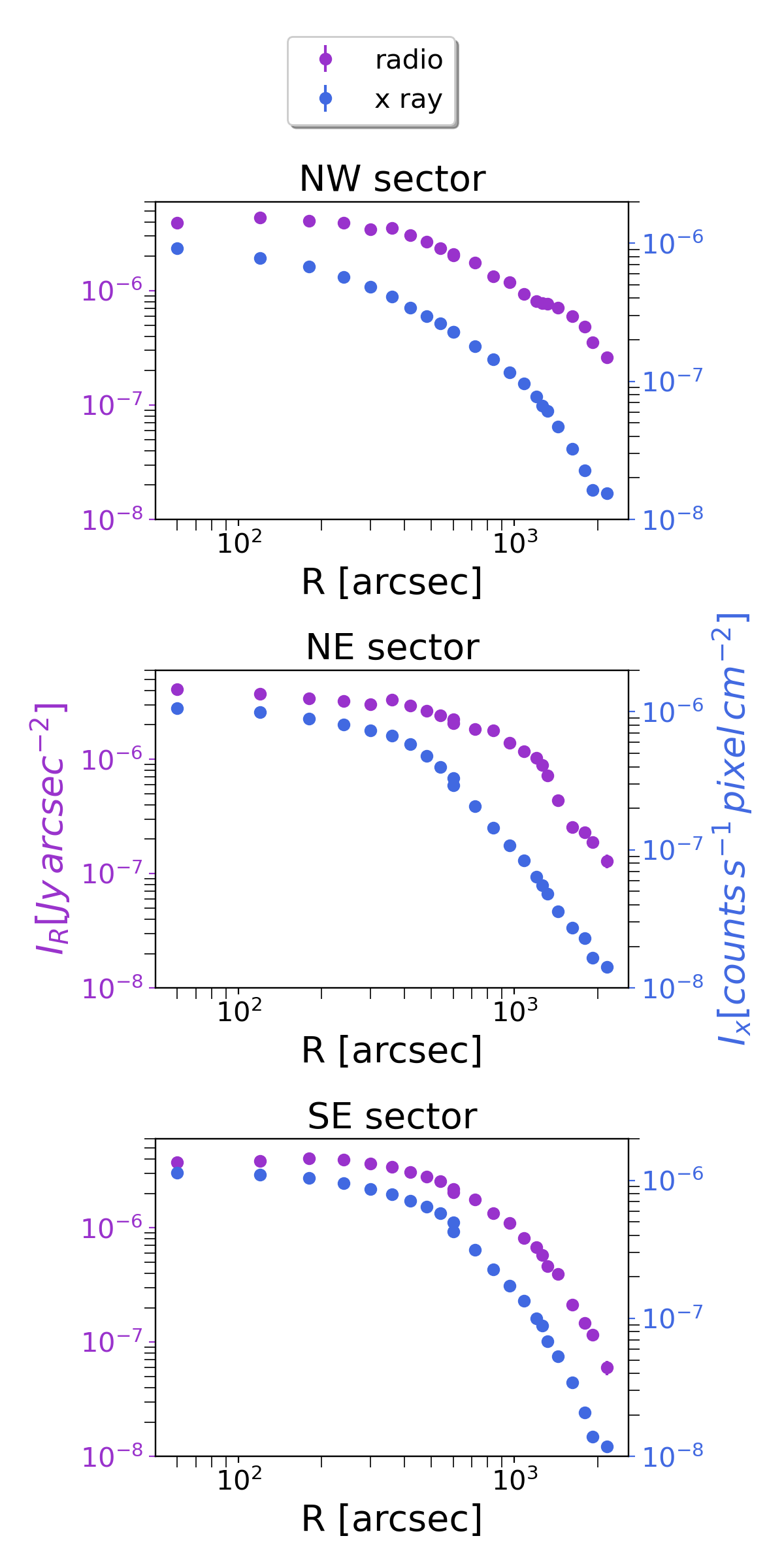}
    \caption{Radial profile of the X-ray brightness and  radio  brightness 
    in the NW, NE, SE sectors (top to bottom). The profiles are computed in elliptical annuli with different width from the centre to the outermost regions as specified in Fig. \ref{fig:halo_profile}, but excluding the SW sector and dividing the remaining area in 3 sectors. The outermost annulus in the NW sector shows contamination from soft X-ray protons and it has been excluded from the following analysis. Statistical errorbars do not appear as they are smaller than the points. }
    \label{fig:plot_profile_radioX}
\end{figure}

\begin{figure}
\vspace{-1pt} 
    \includegraphics[width=0.9\columnwidth]{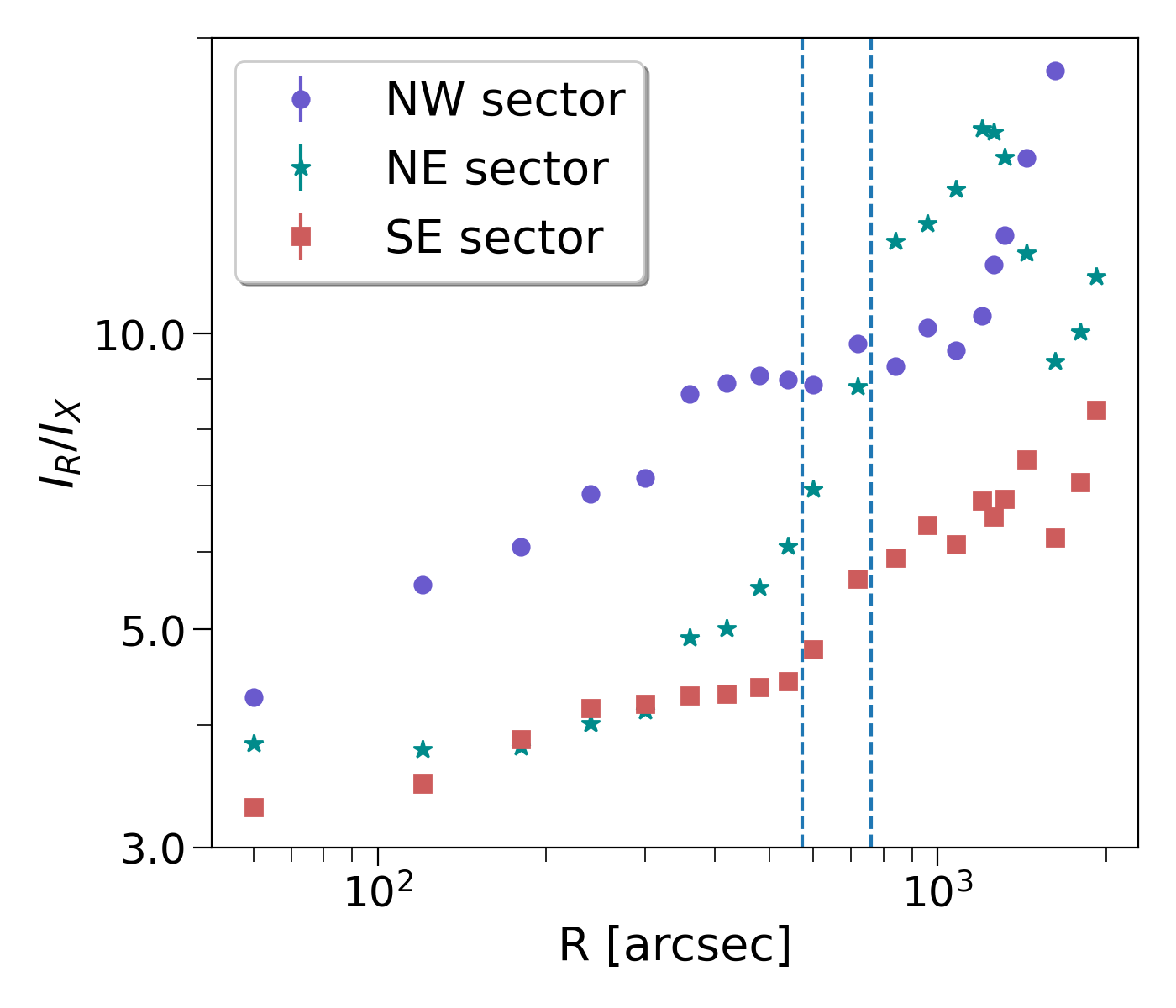}

    \includegraphics[width=0.9\columnwidth]{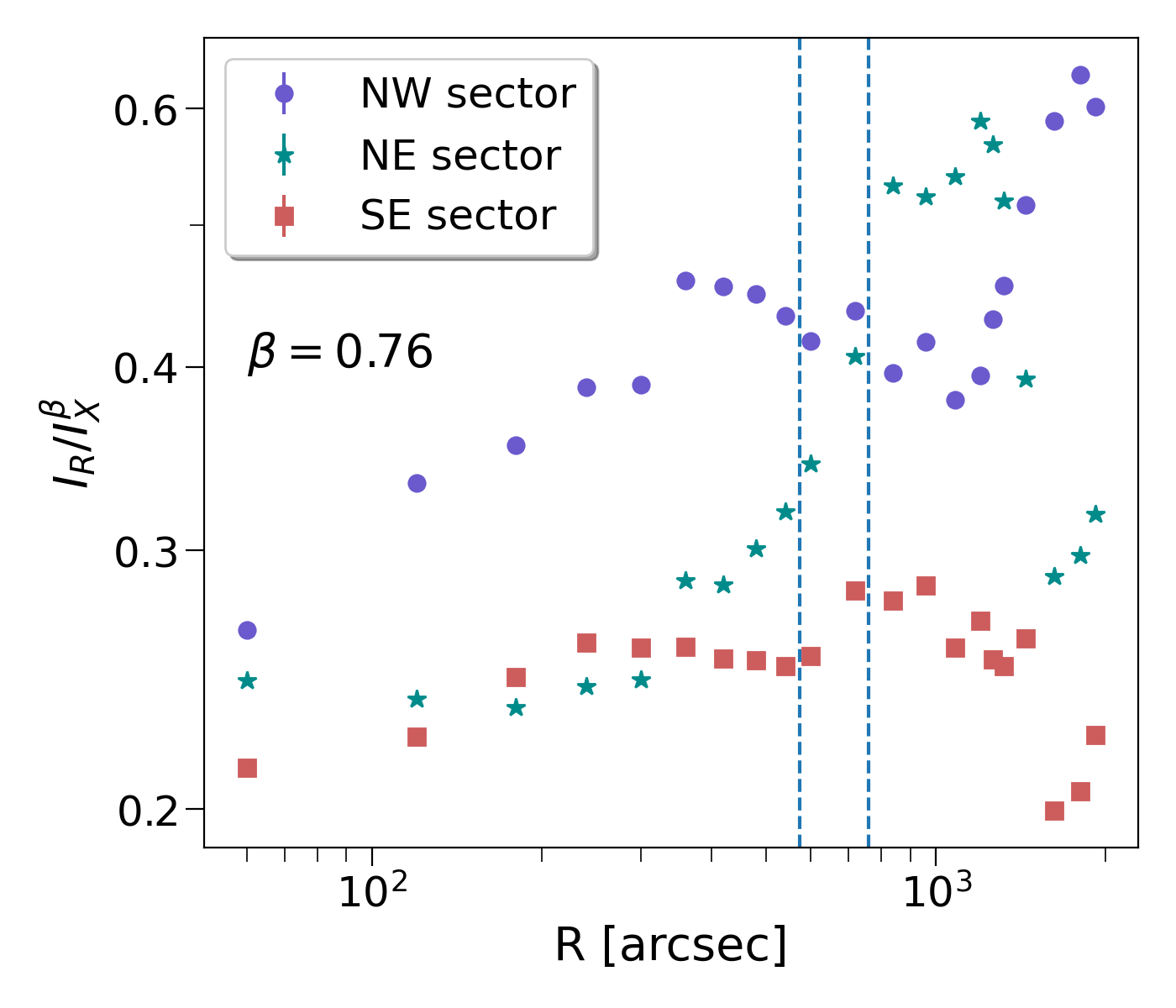}
    
    \includegraphics[width=0.9\columnwidth]{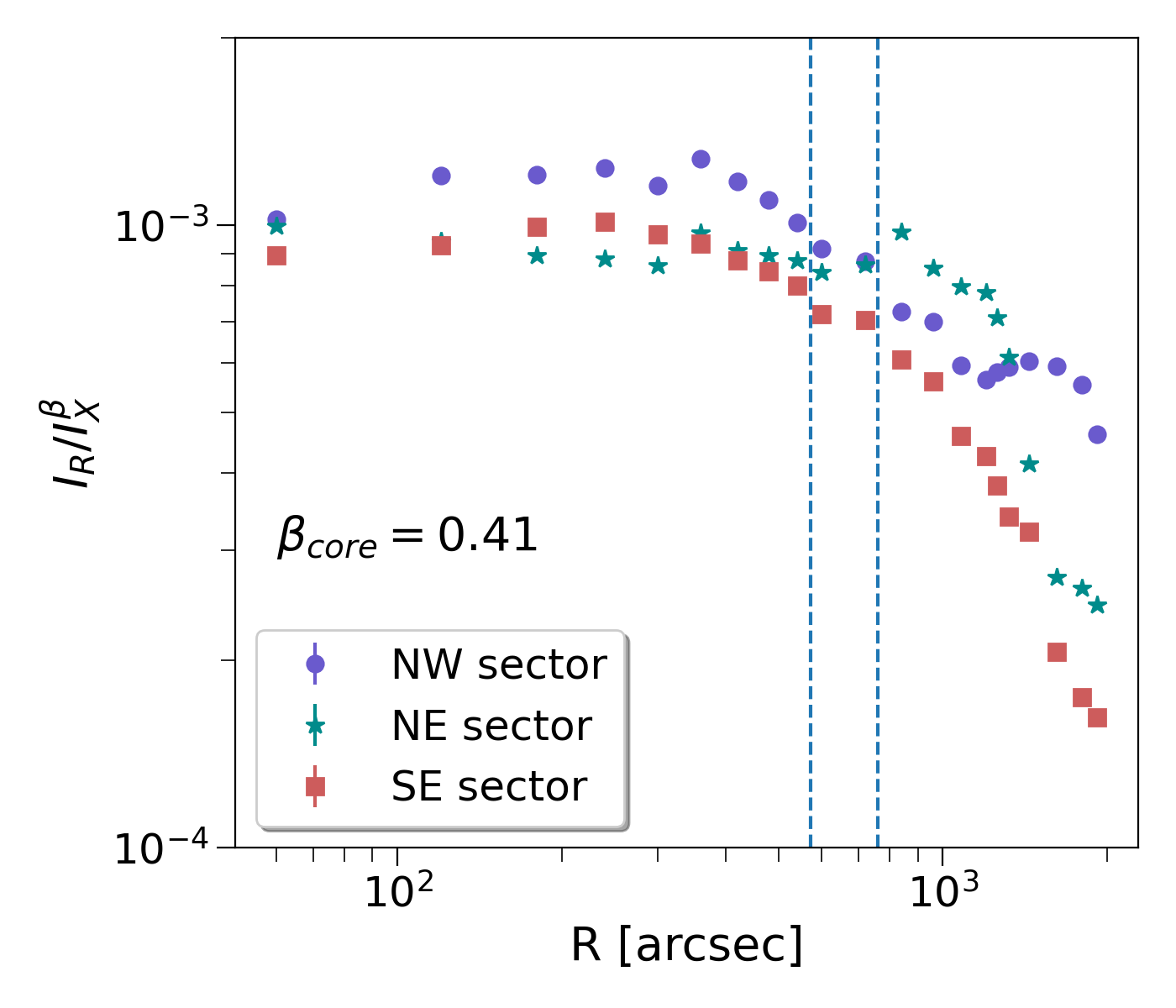}
    
    \caption{Top: Ratio of the radio to X-ray brightness, computed in elliptical annuli as in Fig. \ref{fig:plot_profile_radioX}.  Middle: Ratio of $I_R/I_X^{\beta}$, where $\beta =0.76$ is the best fit correlation coefficient found for the whole halo. Bottom: Ratio of $I_R/I_X^{\beta_{core}}$, where $\beta_{core}=0.39$ is the best fit correlation coefficient found for the halo core. In the three panels, Bullets, stars and squares refer to the NW, NE, and SE sector, respectively. Vertical dashed lines mark the position of $r_1$ and $r_2$, used to separate the halo core from the outer halo. Statistical errorbars do not appear as they are smaller than the points.}
\label{fig:RadioVersusX}
\end{figure}

\subsection{Radial analysis}
The point-to-point analysis above is important to understand the local connection between thermal and non-thermal plasma, and allows one to understand if regions with higher non-thermal energy are traced by high X-ray brightness. The point-to-point analysis has also been used to probe the radial scaling of the radio and X-ray brightness \citep[e.g.][]{Govoni01,Botteon20}.
To better analyse the radial trend of $I_X$ and $I_R$ and to investigate a possible change with radius, we have performed an additional analysis by comparing the X-ray and radio brightness in elliptical annuli and dividing the radio halo in sectors.
Specifically, we have excluded from the analysis the SW sector, where the bridge is, and we have considered separately the NW, NE, and SE sectors. 
The profiles are computed in elliptical annuli with progressively increasing width to gain sensitivity towards low surface brightness emission in the halo outermost regions. X-ray profiles have been computed using the same regions. These plots, shown in Fig. \ref{fig:plot_profile_radioX}, show a different radio profile in the three sectors we have considered, and show clearly a shallower decline of the radio brightness with respect to the X-ray brightness. 
In the NW sector, we detect a sharp decline of the radio brightness at $\sim$1900\asec i.e. the location of the radio front (see sec. \ref{sec:haloFront}). In the NE sector a sharp decline of the radio emission is detected at $\sim$ 1400\asec (670 kpc). This is due to the presence of the filaments of the radio halo (Fig. \ref{fig:zoomHalo}) and to the asymmetry of the halo emission, that is more pronounced towards the W. The radio profile in the SE sector shows, instead, a smooth decline.\\
To inspect the possible change in the relative radial trend of $I_R$ and $I_X$, we show in Fig. \ref{fig:RadioVersusX}, top panel,  the ratio $I_R/I_X$ in Logarithmic base 10 scale, for the different sectors (NW, NE, and SW). The trend is similar in the 3 sectors, with a ratio that is smaller in the central part of the halo and that progressively increases. In the 3 outermost annuli (approximatively at 1600\asec, i.e. $\sim$ 750 kpc ) the ratio decreases in the NE and SE sectors. This plot shows that the ratio of the two quantities is not constant throughout the halo, consistent with the results obtained in the point-to-point analysis.
In the middle panel of Fig. \ref{fig:RadioVersusX}, we show the ratio of $I_R$ to $I_X^{\beta}$, using the value of $\beta$ obtained from the point-to-point analysis ($\beta=0.76$). If the same value of $\beta$ were representative of the whole halo, we would expect to see a horizontal line. Instead, the plot shows that a single value of $\beta$ does not represent the whole halo emission. 
The change in the slope of $\beta$ is shown more clearly in the bottom panel of Fig. \ref{fig:RadioVersusX}, where we plot $I_R/I_X^{\beta_{core}}$, where $\beta_{core}=0.41$ is the best fit value obtained from the point-to-point analysis restricted to the halo core. \\
Hence, we conclude that the $I_R-I_X$ correlation has a slope that changes  with the radial distance from the cluster centre, being flatter in the halo core (where we find $\beta\sim0.41$) and steeper in the outer halo (where we find $\beta \sim 0.76$. While the slope in the core seems to remain constant, the ratio $I_R-I_X^{\beta}$, shown in the middle panel of Fig. \ref{fig:RadioVersusX} indicates a progressive steepening of the correlation with increasing distance from the cluster centre.\\ 

\begin{figure}
\includegraphics[width=0.9\columnwidth]{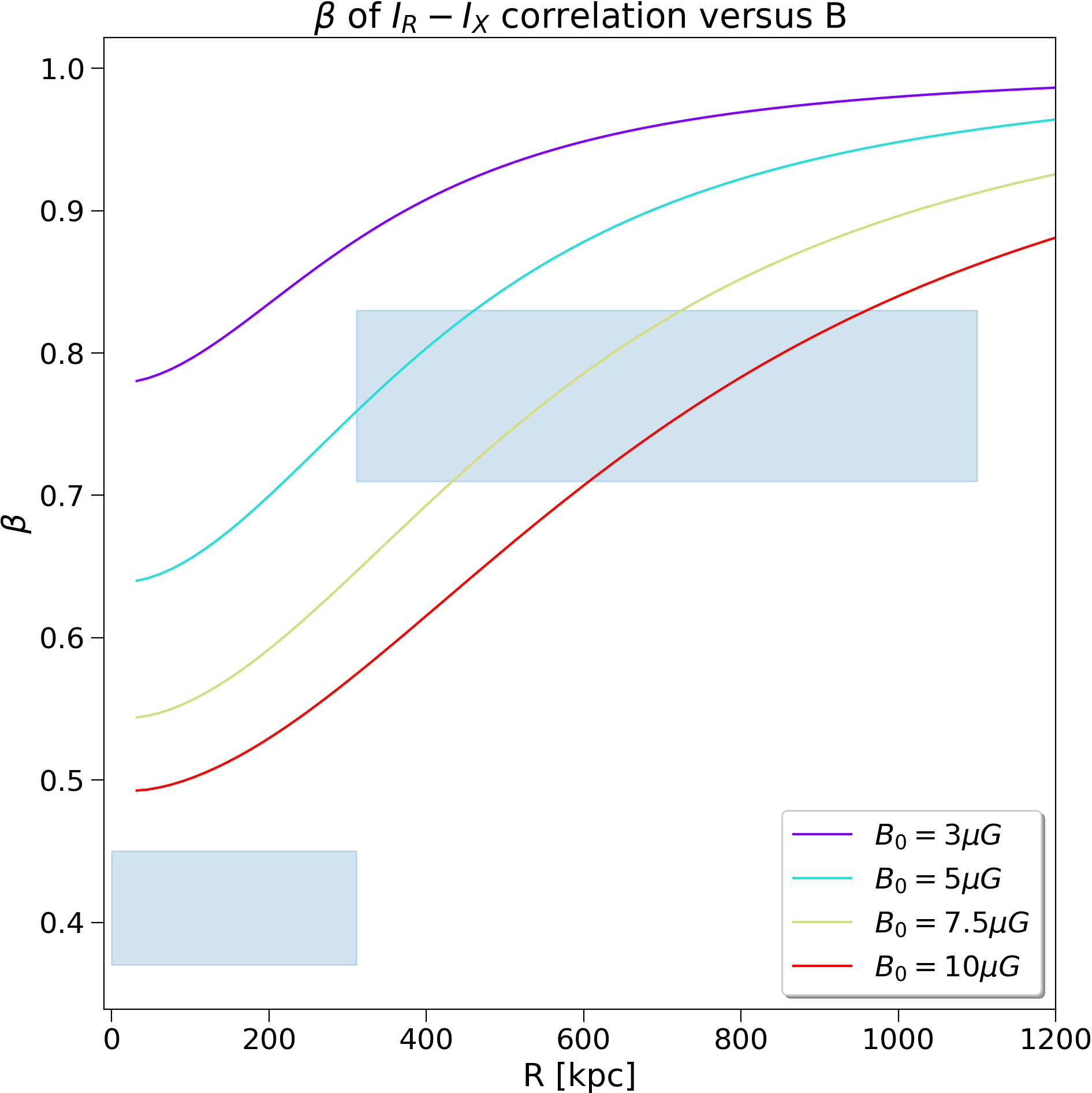}
\caption{Trend of the radio-X ray correlation slope $\beta$ ($I_R=I_X^{\beta}$) with radial distance from the cluster centre. $\beta$ is computed
as the ratio between the logarithm of the theoretical projected radio emissivity to  the logarithm  of the X-ray projected emissivity. The radio emissivity is computed for different values of the central magnetic field $B_0$, from 3 (top curve) to 10 $\mu$G (bottom curve), as specified in the legend. Shaded area refer to the best-fit $\beta$ obtained from data.}
\label{fig:beta_th_core_outer}
\end{figure}

\subsection{Modelling the $I_R-I_X$ correlation and its radial trend}
A steepening of the $I_R-I_X$  correlation outside the core is expected as a consequence of the different relative weight of IC and synchrotron losses in a magnetic field declining with radius. Hence, under some assumptions on the magnetic field profile, we can investigate whether the expected radial drop of magnetic field can be entirely responsible for the steepening of the correlation, or whether additional effects are required.
We assume a magnetic field profile scaling with the thermal gas as $B(r)\propto B_0 \cdot n_e(r)^{0.5}$, consistent with \cite{Bonafede10}. Note that 
in this case the Alfv\'en velocity in the ICM is constant.
The radio emissivity in turbulent re-acceleration models can be expressed as:
\begin{equation}
\label{eq:epsilonr}
    \epsilon_R \propto F \eta_e {{B^2}\over{B^2 +
    B_{IC}^2}}
\end{equation}
where $\eta_e$ is the acceleration efficiency, $B_{IC}$ is the CMB equivalent magnetic field, and  $F$ is the turbulent energy flux:
\begin{equation}
\label{eq:fluxTurb}
F \sim \frac{1}{2} \rho \frac{\sigma_v^3}{ L }
\end{equation}
Here, $\sigma_v$ is the velocity dispersion on scale L, and $\rho$ is the gas density.
Assuming an isotropic distribution of electrons in the momentum space, $f(p)$, the acceleration efficiency is \citep[e.g.][]{BrunettiLazarian07}:
\begin{equation}
    \eta_e \sim F^{-1} \int d^3p {{E}\over{p^2}}
    {{\partial }\over{\partial p}} \big(
    p^2 D_{pp} {{\partial f}\over{\partial p}}
    \big) \approx {{U_{CRe}}\over{F}} ( D_{pp}/p^2 )
\end{equation}
where $U_{CRe}$ is the energy density of reaccelerated electrons, 
and $D_{pp}/p^2$ has different expressions in different re-acceleration models. Specifically,
\begin{equation}
 \frac{D_{pp}}{p^2}  \propto \frac{c_s^2 \mathcal{M}_t^4}{L}
\end{equation}
 in the case of Transit-Time-Damping acceleration with compressive modes \citep{BrunettiLazarian07, Miniati15}, and:
 \begin{equation}
  \frac{D_{pp}}{p^2} \propto \frac{ c_s^3 \mathcal{M}_t^3 }{L v_A}  
 \end{equation}
in case of non-resonant second-order Fermi acceleration with solenoidal modes  \citep{BrunettiLazarian16}. 
If we also assume a constant temperature and a scenario based on a constant turbulent Mach number, the synchrotron emissivity is :
\begin{equation}
    \epsilon_R(r) =\epsilon_R(0) 
    {{X(r)}\over{X(0)}} 
    \left( {{\epsilon_X(r)}\over{\epsilon_X(0)}} \right)^{1 \over 2}
    {{
    1+ \left({{B_{IC}}\over{B_0}}\right)^2 }\over{1+\left({{B_{IC}}\over{B(r)}}\right) ^2} \left({{\epsilon_X(0)}\over{\epsilon_X(r)}}\right)^{1 \over 2} }
\end{equation}
where $\epsilon_X \propto n^2$ is the X-ray (bremsstrahlung) emissivity and 
$X= \frac{U_{CRe}}{U_{th}}$ the ratio of the energy density of CRe to thermal gas.
From the radio and X-ray emissivity, we have computed the projected radio brightness assuming constant $X$, and  different values of the central magnetic field, $B_0$, ranging from 3 $\mu G$ to 10 $\mu G$, and the X-ray brightness using the parameters obtained by \citet{Briel01}. Then, we have computed the slope of the correlation $I_X-I_R$ as a function of the radial distance from the cluster centre. The trend of $\beta$ versus the radial distance is shown in Fig. \ref{fig:beta_th_core_outer} for different values of $B_0$. We have overplotted  as shaded areas the values obtained in the halo core and in the outer halo. \\
Although all the models predict a steepening of the radio-X-ray scaling with radius, none of them is able to reproduce all observed values. In particular, except for the profile with $B_0=5\mu$G, the magnetic field profiles that would be compatible with the value of $\beta$ in the outer halo are higher than those derived from RM studies \citep{Bonafede10}. 
A possible scenario to explain the observed trend, is to assume that the ratio of the CRe to thermal gas energy densities ($X$)  increases with  radial distance from the cluster centre. This is indeed expected, as the lifetime of CRe in the ICM increases with the distance from the cluster center due to the lower Coulomb and synchrotron losses. However, it is the first time that a radially decreasing $X$ seems to be suggested by observational data. \\
As for the calculations done in Sec. \ref{sec:haloSpectrum_theo}, we stress that the calculations outlined here are subject to several assumptions and the decrease of $X$ with the distance from the cluster centre is only one of the possible causes of the observed $I_R-I_X$ slope.

%

\begin{figure}
\centering
\includegraphics[width=\columnwidth]{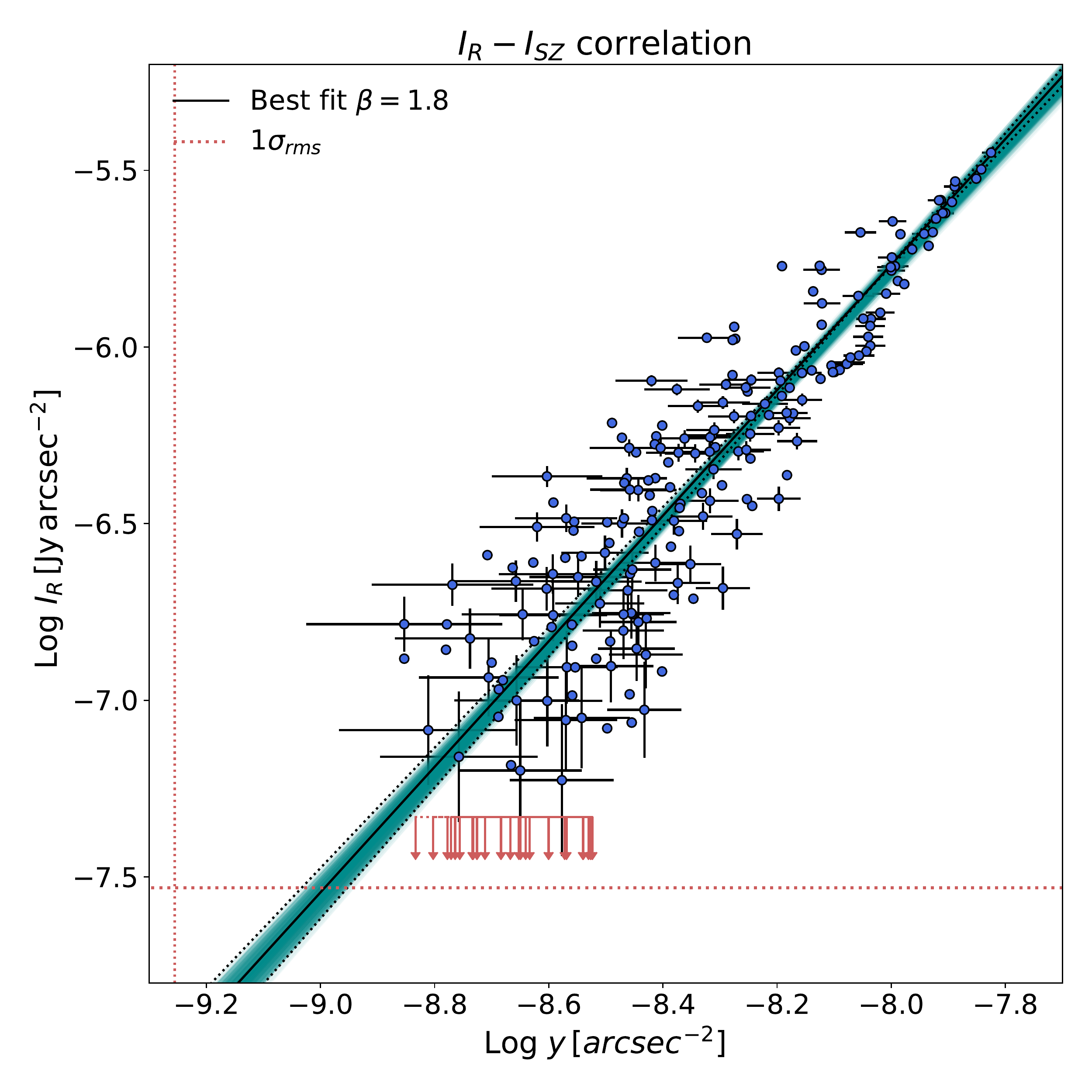}

\caption{Radio-SZ correlation for the radio halo. The radio image has been convolved to a resolution of 5$'$ and the source NGC4839 has been masked. The errorbars refer to the statistical errors of the two quantities. Arrows mark the $2\sigma_{\rm rms}$ upper limits. The vertical and horizontal dotted lines  marks $1\sigma_{\rm rms}$ for the $y$-parameter and radio image, respectively. Errorbars are plotted every second value.}
\label{fig:corr_radioSZ}
\end{figure}

\section{Radio - thermal pressure correlation in the radio halo}
\label{sec:radioSZ}
Resolved SZ maps of the Coma cluster \citep{PlanckComa} can be used to understand the connection between the thermal gas and the radio emission.
Since the comptonization parameter $y$ is proportional to the gas pressure integrated along the line of sight ($y \propto n_e T$),  it is less contaminated by cold gas clumps with respect to X-ray emission.
We have computed the $I_{SZ} - I_R$ point to point correlation, following the same approach as described above for the $I_R - I_X$ correlation. We have used the radio image convolved to 5 arcmin resolution, to match the $y$-map resolution.\\
Initially, we have fit $I_{SZ}$ versus $I_R$ in the same region as the one considered in the $I_R-I_X$ correlation, finding a super-linear slope
$\beta_{SZ}= 1.76\pm0.08$. 
Assuming an isothermal model, one would expect that the value found from the $I_{SZ} - I_R$ correlation would be twice the scaling of the $I_R-I_X$. Hence, given the values found from the $I_R-I_X$ correlation, one would expect $\beta_{SZ}= 1.4 - 1.6$. There is a small tension between SZ and X-ray, that could be due e.g. to dense and cold X-ray clumps in regions of low surface brightness emission. Understanding the details of this small tension is not the scope of this work, what is relevant for our analysis is that regions of high non-thermal energy are regions of high thermal energy, as probed by the  $I_{SZ} - I_R$ correlation. \\
The $y$-map is more extended than the \emph{XMM-Newton} mosaic, and covers regions
at the cluster outskirts where we have a radio halo detection. Hence, we can use 
the $y$ parameter map  to investigate the correlation between thermal and non-thermal regions up to larger radii from the cluster centre, though we miss the resolution given by X-ray data.
We have fitted $I_{SZ}$ versus $I_R$ out to a distance of $\sim$1.3 Mpc ($\sim$2800\asec radius) from the cluster centre, that is the maximum distance where we have detected the radio halo. As shown in sec. \ref{sec:radioXcorrelation}, we have indication from the X-ray analysis that the slope would further increase when outer regions of the halo are considered. We find $\beta=1.78 \pm 0.08$, which is slightly steeper than the slope found in the inner 2400\asec radius, though consistent within the errors. The $I_{SZ} - I_R$ correlation is shown in Fig. \ref{fig:corr_radioSZ}. Both X-ray and SZ analysis show that when outer regions of the halo are considered,  the correlation slope increases, i.e. the outermost regions of the radio halo have a different ratio of thermal/non-thermal energy than the inner ones.\\
A correlation between the radio emission and the $y$-signal has been firstly obtained by \cite{PlanckComa}, who fitted $I_{SZ}$ versus $I_R$, finding a quasi-linear relation: $I_{SZ} \propto I_{R}^{0.92 \pm 0.04}$, that would correspond to our $\beta \sim 1.1$. They used the WSRT 325\,MHz map and  $y$  images at 10 arcmin resolution, and extracted the radio and $y$-signal from  $r<50$ arcmin region.
As we find a steeper slope, using a lower frequency radio image we investigate in the following the possible causes of the different trend.
First of all, we are looking at the $I_R - I_{SZ}$ correlation using a radio image at a different frequency than used in \citet{PlanckComa}. 
In addition, we are able to perform a more accurate subtraction of the contaminating sources, as the highest LOFAR resolution is 6\asec, and we are less affected from calibration artefacts than the WSRT image.  Residuals from contaminating sources would increase the values of the radio brightness in each box, and this effect could be more prominent for boxes at the halo periphery. In this case, the effect would be a flattening of the correlation.
The data we have at the moment of writing do not allow to exclude that this is the cause of the different slope found with LOFAR and WSRT. However, we note that a higher value of the correlation slope for the $I_R - I_X $ at lower frequencies has recently been found by Rajpurohit et al. (submitted) in the halo of the cluster Abell 2256, and by \cite{Hoang21} in the halo of CIG 0217+70.

\begin{figure*}
\centering
\includegraphics[width=2\columnwidth]{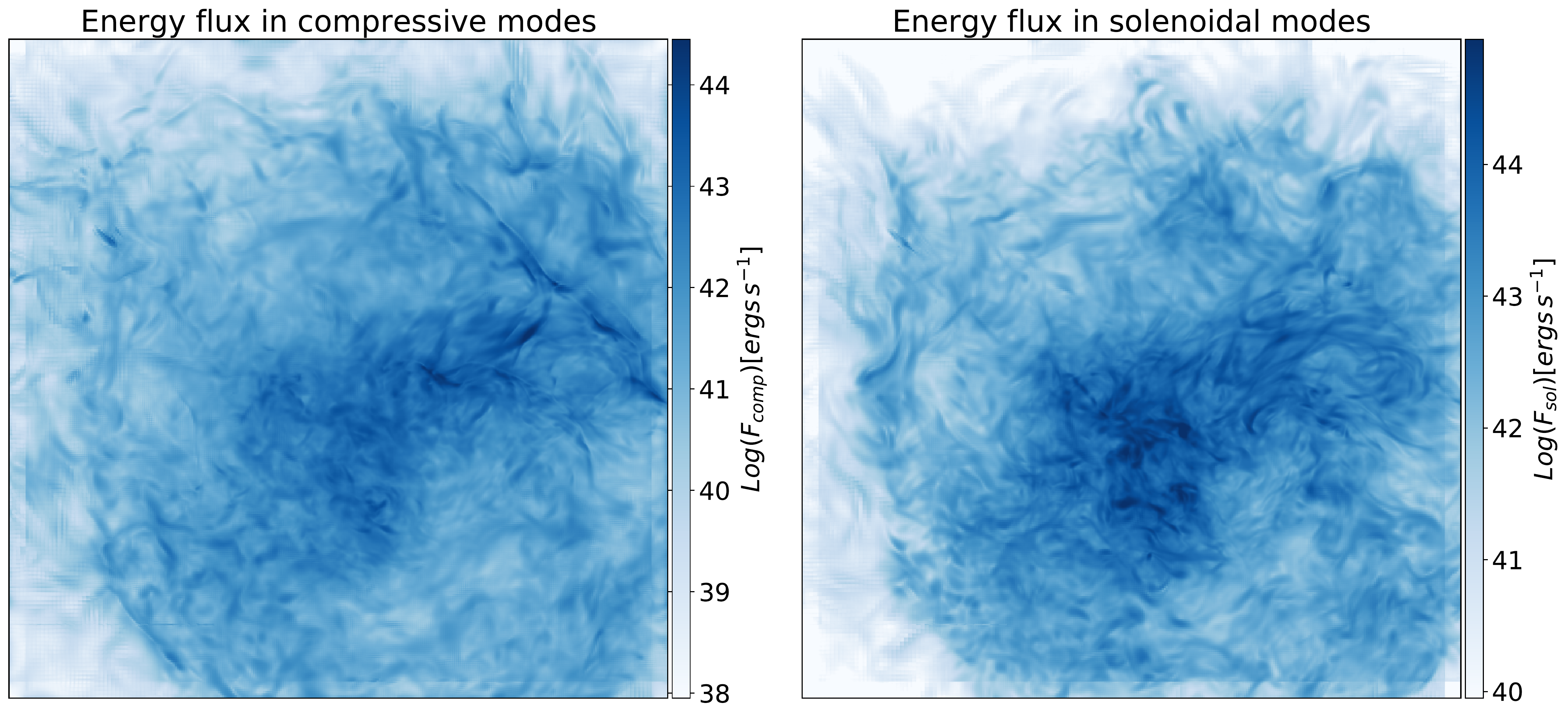}
\caption{Projected $F_{comp}$ (left) and  $F_{sol}$ (right) from MHD cosmological simulations of a Coma-like cluster, from \citet{va18mhd}. Each map has a size of 2560 kpc.}
\label{fig:sim_images}
\end{figure*}

\begin{figure*}
\centering
\includegraphics[width=\columnwidth]{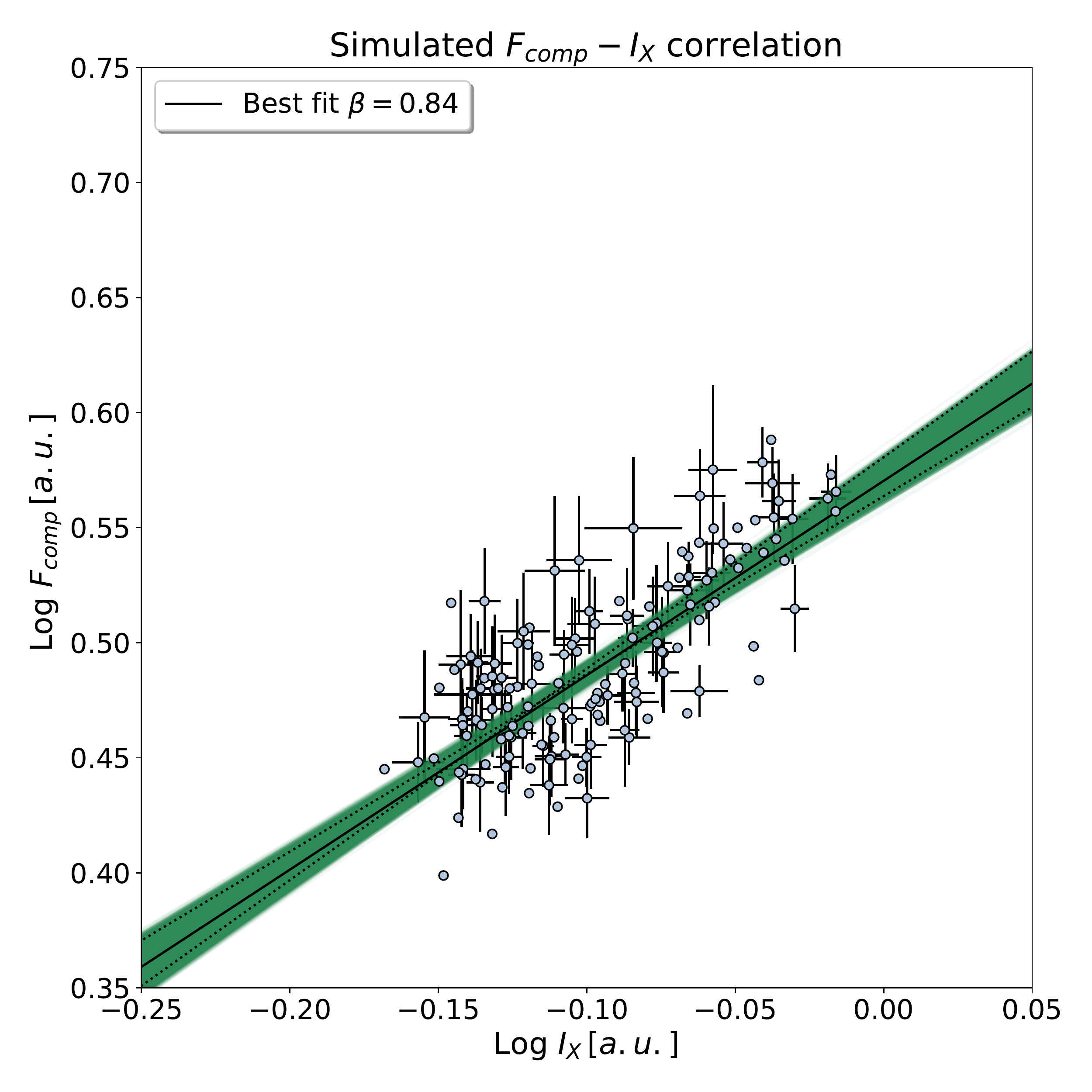}
\includegraphics[width=\columnwidth]{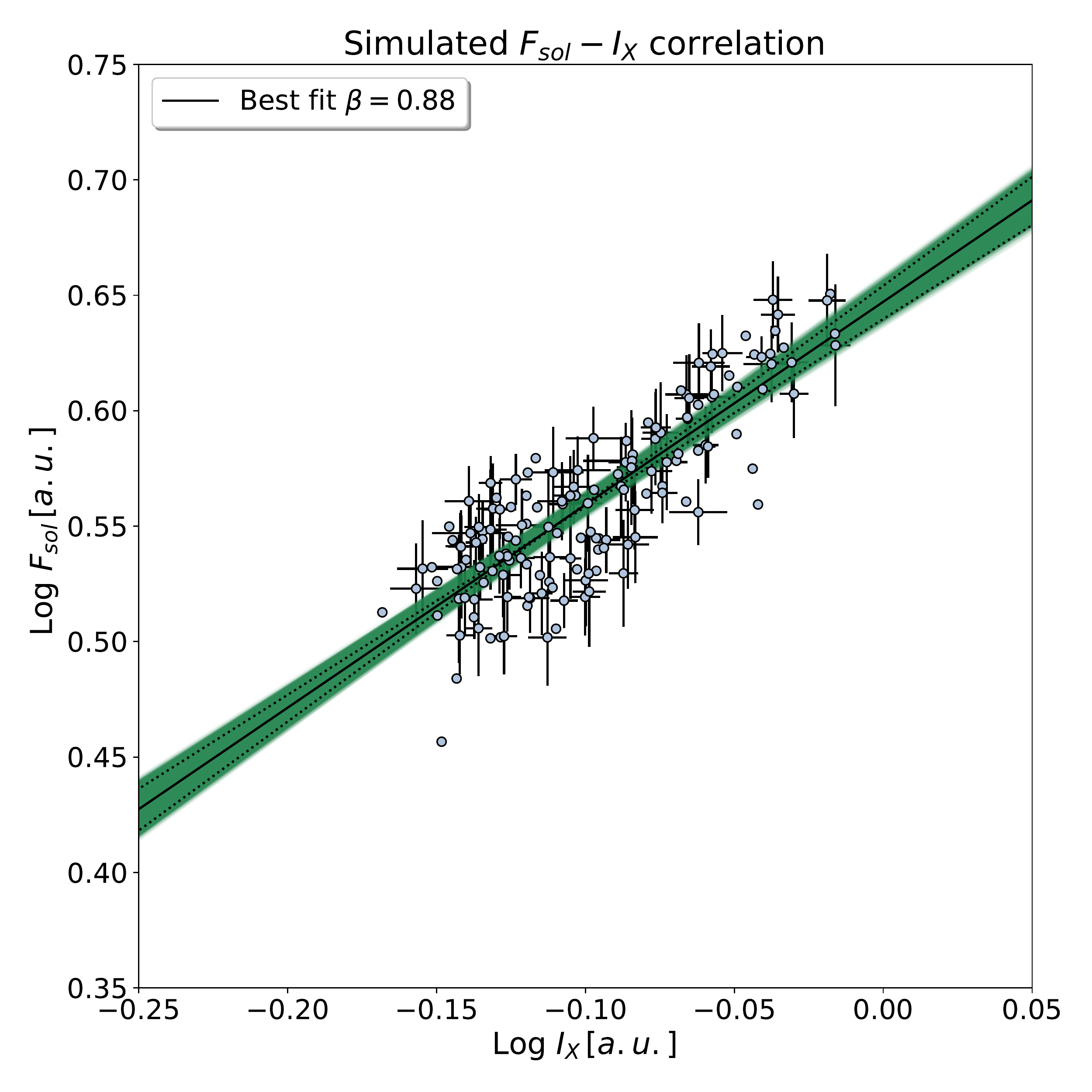}
\caption{Left panel: $F_{Comp} - I_X$ correlation from simulated data. Right panel: $F_{Sol}- I_X$correlation from simulated data.}
\label{fig:corr_sim}
\end{figure*}

\section{Comparison with numerical simulations}
\label{sec:simulations}
Having the radio halo resolved in great detail, we can try to understand its origin with the help of numerical simulations. Specifically, we try in this Section to understand whether the observed scaling  $I_R - I_X$ can provide useful information to better constrain the particle acceleration mechanism.

\subsubsection{Simulated thermal to non-thermal correlations}
We use a  Coma-like  galaxy cluster, simulated at high resolution and with ideal Magneto-Hydrodynamics using the cosmological Eulerian code ENZO (enzo-project.org) by \citet{Vazza18}. This system has a $z \approx 0.02$ total mass comparable with the  real Coma cluster, and it shows a radial decline of the magnetic field compatible with the estimates from  Rotation Measures \citep{Bonafede10,va18mhd}. The simulation includes  eight levels of Adaptive Mesh Refinement (AMR) to  increase the spatial and force resolution in most of the innermost cluster volume, down  to  $\Delta x=3.95 \rm ~ kpc/cell$. While this simulation assumes 
an initial volume-filling background of weak magnetic field, $B_0=10^{-4} \rm \mu G$ (comoving) at $z=40$, the low redshift properties of the magnetic field are fairly independent of the exact origin scenario, due to the effect of the efficient small-scale dynamo amplification \citep[e.g.][]{2009MNRAS.392.1008D,va21b}.
Using the filtering technique described in \citet{va17turb}, we have computed the turbulent energy flux, $F_{Comp,Sol}$, associated to the compressive and solenoidal velocity components: 
\begin{equation}
   F_{Comp,Sol}= \rho \frac{\sigma_{v_{Comp,Sol}}^3}{L}\times \frac{B^2}{B^2 + B_{IC}^2},
\end{equation}
where, similarly as in Eq. \ref{eq:fluxTurb}, $\sigma_{v_{Comp,Sol}}$ is the dispersion of the compressive, solenoidal velocity field on scale $L$, different for solenoidal and compressive modes.
From $F$, one can compute the simulated synchrotron luminosity (see Eq. \ref{eq:epsilonr}) as:
\begin{equation}
    L_{Comp,Sol}= \eta_e F_{Comp,Sol}
\end{equation}
The constant $\eta_e \leq 1$ gives the dissipation efficiencies for solenoidal and compressive modes into cosmic ray acceleration, which depend on the complex physics of cosmic ray acceleration via Fermi II process \citep[see e.g.][]{Miniati15,BrunettiLazarian16,BV20}. However, since in this application to our new LOFAR observations of Coma we are only concerned in the relative distribution of the two energy fluxes, we fix $\eta$ for simplicity, acknowledging that both fluxes will represent an overestimate (likely by a factor $\sim \geq 10^2$) of the effective dissipation onto cosmic ray acceleration. Hence, we have used $F_{Comp,Sol}$ as a proxy for the synchrotron luminosity instead of $L_{Comp,Sol}$.
The maps of projected  $F_{Comp}$ and $F_{Sol}$ are shown in Fig. \ref{fig:sim_images}. 
We have computed the correlation between $F_{Comp,Sol}$ and the simulated X-ray brightness $I_X$ as we have done for the radio and X-ray emission.
We note that some assumptions must be done to compare the simulated $F_{Comp,Sol}$ to the observed radio emission. Specifically, we have assumed that the halo emission is generated by turbulent re-acceleration, and  that $\eta_e$ is constant throughout the cluster volume. Furthermore, as we are not modeling the CRe component, we have averaged $F_{Comp,Sol}$ on scales larger than the electron diffusion length at 144 MHz.
The diffusion length $l_e$ of electrons emitting at 144 MHz can be estimated at first order as 
\begin{equation}
    l_e \sim 2 \sqrt{D \tau_{rad}} \sim 100 \rm kpc.
\end{equation}
Here, $\tau_{rad}$ is the electrons' radiative age at 144 MHz ($\tau_{rad} \approx 200$ Myr) and 
$D$ is the spatial diffusion coefficient. A simple estimate for $D$ (i.e. ignoring pitch angle scattering along magnetic field lines) can be obtained from the typical scale for MHD turbulence in the ICM ($l_A \sim 0.1-0.5$ kpc) as $D \approx \frac{1}{3} l_A c$.\\
We have computed the mean of  $F_{Comp,Sol}$ and $I_X$  in boxes of 160 kpc side, that is larger than $l_e$ and comparable to the size of the boxes used to compute the $I_R-I_X$ correlation from data (see Sec. \ref{sec:radioXcorrelation}). \\
In Fig. \ref{fig:corr_sim}, the two correlations obtained with simulated data are shown. Both $F_{Comp}$ and $F_{Sol}$ are positively correlated with the simulated $I_X$. The correlation slope is similar, though a bit steeper in the case of  $F_{Sol}$-$I_X$ (see Tab. \ref{tab:corr_sim}. Both slopes are  consistent with the observed $I_R - I_X$ correlation, supporting the connection between turbulence and radio diffuse emission. We note that the power in the solenoidal energy flux is $\sim$ 10\% higher than in the compressive energy flux. Once the CRe emission is properly modelled, this could be used to disentangle the role played by the two modes for CRe re-acceleration.

\subsection{Correlations in the centre and peripheral regions}
As shown in the previous section, the global $I_R-I_X$ in the Coma cluster can be recovered using both compressive and solenoidal energy fluxes as a tracer for the radio emission. 
MHD simulations show that solenoidal modes are dominant in the cluster central regions, and compressive modes dominate the cluster outskirts \citep{Miniati15,Vazza17}. We have seen in Sec. \ref{sec:radioXcorrelation}, that the different $\beta$ obtained in the halo core and in the outer halo can not be explained by the decline of the magnetic field only. Here, we investigate whether different turbulent modes could play a role. We have fitted the simulated $F_{comp,sol}$ versus the simulated X-ray brightness $I_X$ in the cluster core and in the cluster external regions, separately. We used boxes of 32 kpc size to have a better sampling of the halo core.
The results are listed in Table \ref{tab:corr_sim}. The best-fit slope in the halo core is flatter than in the outer halo for both solenoidal and turbulent energy flux. The Pearson correlation coefficient, though, indicates that the correlation between $I_X$ and $F_{comp,sol}$ is weak in the halo core. \\
Hence, we conclude that globally the trends are in agreement with the scenario where turbulent re-acceleration produces the radio halo. In order to understand the process in more detail, the CRe distribution needs to be modelled.

\begin{table}[]
 \caption{Correlation between thermal and non-thermal simulated quantities}
    \centering
    \begin{tabular}{c c c c}
    \hline
    \hline 
     & $\beta$  & 10\% -- 90\%  & $\rho_{P}$\\
 $F_{Sol}  -I_X$   & 0.88 & 0.94 -- 0.81 &  0.8\\
$F_{Comp} - I_X $   & 0.85 & 0.92 -- 0.77  &  0.8\\ 
 \hline 
 \multicolumn{4}{c}{Halo core} \\
 $F_{sol} - I_X $  & 0.5 &0.6 - 0.4 &  0.2\\
 $F_{Comp} - I_X $ & 0.5 &0.6 - 0.3 &  0.3\\
 \hline
  \multicolumn{4}{c}{Outer halo} \\
 $F_{sol} - I_X $  & 0.83 &0.91 - 0.75 &  0.8\\
 $F_{Comp} - I_X $ & 0.78 &0.87 - 0.69 &  0.8\\

\hline \hline
    \end{tabular}

    \label{tab:corr_sim}
\end{table}

\begin{figure*}
\centering
\includegraphics[width=\columnwidth]{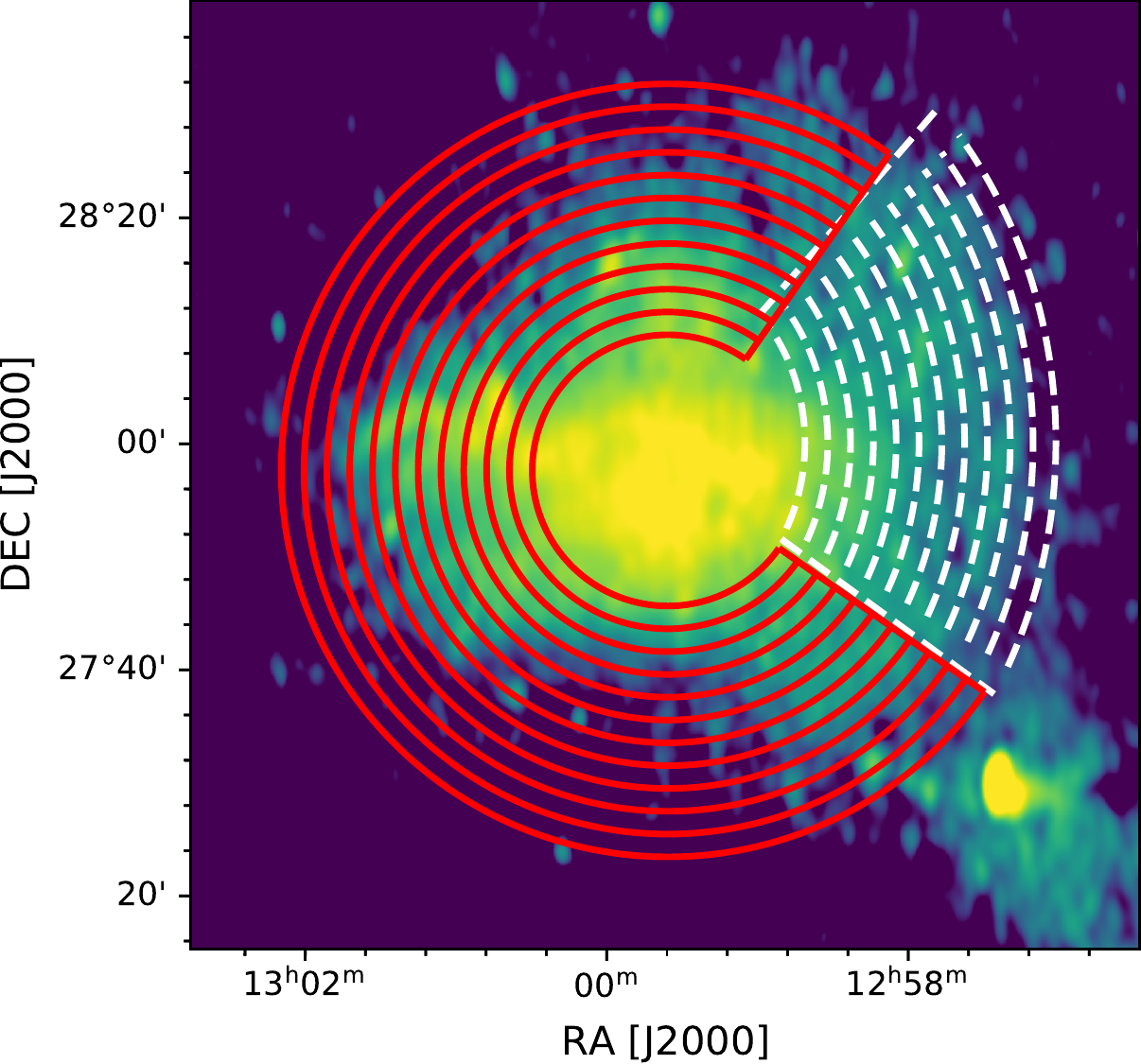}
\includegraphics[width=\columnwidth]{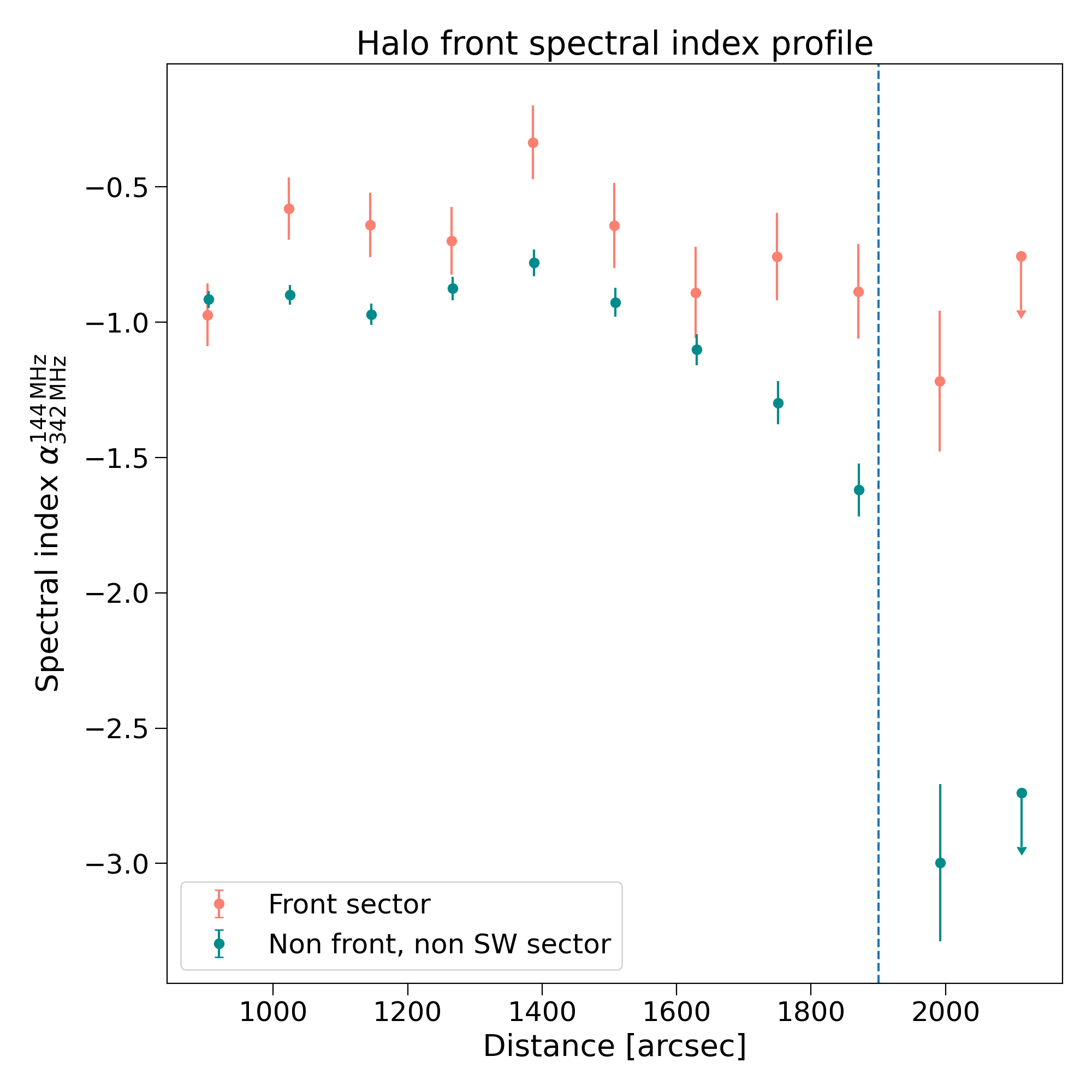}
\caption{Left panel: annuli used to compute the spectral index trend in the halo front region (white - dashed) and in the outer halo (red). Right: Spectral index trend computed in the annuli shown in the left panel. Arrows indicate 3$\sigma$ upper limits. The vertical dashed line indicates the position of the halo front. 
Errorbars only show statistical errors,  the values of the spectral index are also affected by the flux calibration uncertainties of WSRT and LOFAR (10\% and 15\%, respectively) that would contribute with an additional error of 0.2.}
\label{fig:spix_front}
\end{figure*}

\section{The halo front}
\label{sec:haloFront}
Previous studies  pointed out the presence of discontinuities in the thermal gas cluster properties located at the western edge of the radio halo \citep{Markevitch10,PlanckComa,Simionescu13}. This discontinuity, detected in both the Comptonisation parameter $y$ and in the temperature and deprojected density profiles, is consistent with an outwardly-propagating shock front, located at $\sim$ 33 arcmin (910 kpc) from the cluster centre. More recently, \citet{Churazov21} used X-ray data from eROSITA and confirmed the presence of a shock wave in the West, with a Mach number  $M \sim 1.5$. This shock is interpreted as a secondary shock, or ``mini accretion shock" driven by the first passage of the NGC 4839 group through the cluster before reaching the apocenter and inverting its orbital motion. According to this scenario, both the relic and the W shock would be caused by the merger of NGC 4839 with the Coma cluster. During its first passage NGC4839 would have driven a first shock that should now be located at the position marked by the radio relic. The gas displaced by the merger would settle back into hydrostatic equilibruim, forming a ``mini accretion" shock. 
We refer the reader to  \cite{Burns94}, \cite{Lyskova19}, \cite{Zhang21}, \cite{Churazov21} for a more detailed explanation of the proposed merging scenario.\\
Our LOFAR image confirms the presence of an edge of the radio halo towards the West (see Fig. \ref{fig:profile_front}), that could be in line with the scenario explained above.
Shocks in the ICM are often associated with radio relics, where they leave a clear imprint on the spectral index distribution.
The spectral index profile is flatter where particles are freshly re-accelerated and steeper moving towards the downstream region where particles radiate their energy via synchrotron and Inverse Compton losses. We have investigated whether a similar trend is found in the ``halo front", computing the spectral  index profile in annuli that follow the front  (Fig. \ref{fig:spix_front}). To derive the spectral index values, we have used the same images presented in Sec. \ref{sec:haloSpectrum}\footnote{we recompute the spectral index radial trend in this section to better highlight the differences between the front region and the rest of the halo, while in Fig. 9 we have divided the halo in 4 identical sectors. For the same reason, the spectral index is computed out to a smaller distance from the cluster centre than in Fig. 9.}. 
In Fig. \ref{fig:spix_front}, we show the spectral index profile obtained in annuli  that follow the front, in comparison with the global spectral index trend obtained in annuli centred on the cluster and excluding the halo front region (Fig. \ref{fig:spix_front}, left panel). 
We note that in the direction of the front, the spectral index is flatter, but no steepening in the putative downstream region is present. However, the  outermost annuli seem to follow a different trend than the global halo profile: the spectral index remains almost constant, while a radial steepening is detected when the whole halo is considered. It is possible that we are limited by the resolution, as the width of the annuli is $\sim$2$'$, corresponding to $\sim$60 kpc, that is larger than the electron cooling length in the post shock region \citep[e.g.][]{KangRyu12}.
Hence, it is possible that we do not have the resolution to separate the shock front from the post-shock region, where particles have already experienced strong radiative losses.
An alternative scenario is that the radio front could be radio plasma re-accelerated by the same mechanism responsible for the halo emission, and then dragged by the shock passage and compressed by it. 
Possibly, future high-resolution observations at higher or lower frequency will allow us to understand whether the halo front shows the typical shocked
spectral index profile that we detect in radio relics or not.\\

\begin{figure}
\centering
\includegraphics[width=\columnwidth]{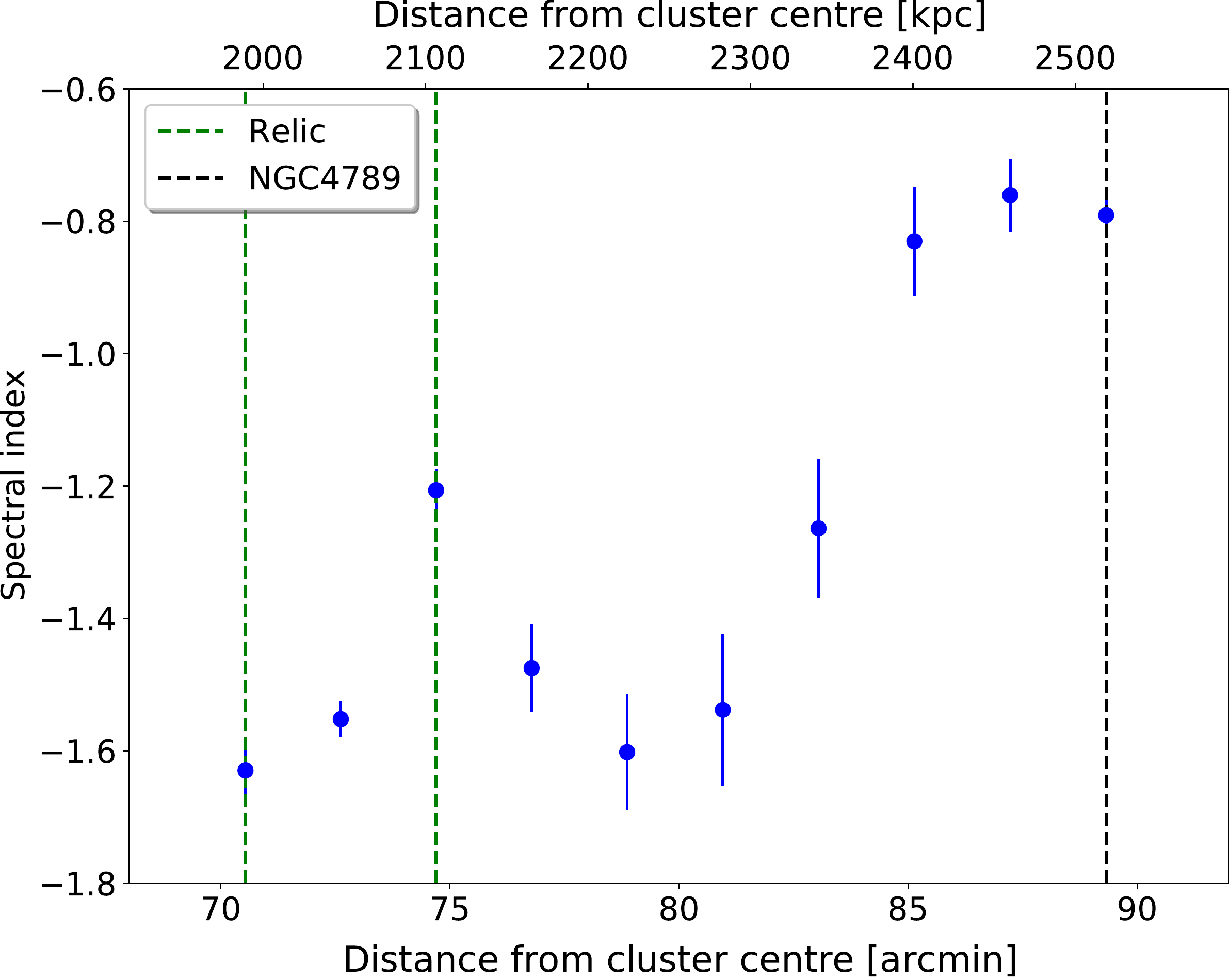}
\caption{
Spectral index profile between 326 MHz and 144 MHz of the emission called relic-NAT connection. Images at the WSRT resolution (100\asec $ \times $ 150\asec) have been used to compute the spectral index in each box.}
\label{fig:Relic_NAT_profiles}
\end{figure}


\section{The relic-NAT connection}
\label{sec:relicNAT}

The diffuse emission connecting the relic to the NAT radio galaxy NGC4789 is imaged here with unprecedented resolution, that allows us to detect substructures in its surface brightness distribution. 
In particular, the bent jets of  NGC4789 do not blend smoothly into the diffuse emission, as observed in other cases of radio galaxies nearby relics \citep{Bonafede14b, vanWeeren17_Nature, Stuardi19}, and two discontinuities between the endpoint of the jets and the diffuse emissions are detected. 
At the centre of the diffuse source region, there is a bright transverse bar.
Such a bar has been detected in other bent tails (e.g. in Abell 2443, \citealt{Clarke13}, and recently in the Shapley supercluster \citealt{Venturi22}) and are predicted by simulations of interacting AGN tails and shocks \citep{Nolting19}. 
The length of this source, measured from the relic's edge down to the endpoints of the jets, is 10 arcmin, corresponding to 280 kpc at the Coma redshift.\\ 
To investigate the possible connection of the radio plasma with the thermal gas, we have used the ROSAT image and the radio image at 35 \asec resolution, and investigated the existence of a correlation between the thermal and non-thermal plasma, as done already for the radio halo (see Sec. \ref{sec:radioXcorrelation}), and for the Coma bridge \citep{Bonafede21}. 

The average brightness profile of the relic-NAT connection  is largely constant, increasing close to the relic edge where the radio bar is located. In addition, the low counts in the X-ray image do not allow us to make a proper analysis. Hence, though no correlation or anti-correlation between the two quantities seems to be present, no firm conclusions can be derived, and we can not rule out that the radio emisison originates from phenomena similar to those responsible for the bridge. 
However, given its morphology, we will investigate below an alternative 
scenario.\\
The spectral index trend along the Relic-NAT connection main axis is shown in Fig. \ref{fig:Relic_NAT_profiles}. We have used the WSRT R image (see Tab. \ref{tab:images}, \citealt{Giovannini91}) that has been used already in \citet{Bonafede21} to analyse the bridge. Although the higher resolution of the image by \cite{BrownRudnick11} would provide a better description of the spectral index trend, calibration errors from ComaA are strongly affecting that region. 
The spectrum of the radio emission computed between 144 MHz and 326 MHz shows a gradual steepening from regions close to the AGN core towards the relic's outer edge, reaching values of $\alpha = -1.6 \pm 0.2$. 
At the relic's outer edge, the spectral index flattens to $\alpha=-1.2 \pm 0.2$, and it steepens towards the NE, i.e. the putative post-shock region, reaching again $\alpha= -1.6 \pm 0.2$. We note that the spectrum starts to flatten already in front
of the relic edge (at $\sim$77$^{\prime}$ from the cluster centre), though within the error that value is consistent with the steepest point. This apparent flattening could be due to projection effects, as if the relic has a velocity component along the line of sight, some relic emission could appear in front of the relic edge in projection.
\\
Overall, the spectral trend detected along the tail of NCG4789 and in the relic-NAT connection is consistent with AGN particle ageing. NGC4789 would be moving towards the SW of the cluster injecting particles into the ICM. Thus more recently injected particles are closer to the AGN core than the older ones that have been left behind. The shape of the source and the connection with the radio relic make this source peculiar and suggest a link between the relic-NAT connection and the shock wave that would power the radio relic.\\
However, according to the merging scenario outlined in several papers \citep{Venturi90,Giovannini91,Churazov21}, NGC4789 would be in the pre-shock region. Hence, it is puzzling to understand how features like the bar could have been formed, because that emission has not yet interacted with the shock.
\cite{EnsslinGK01} have proposed that the plasma from NGC4789 is dragged by the infalling matter (falling into Coma's cluster) to the location of the relic, where it is re-energised adiabatically by the shock wave.\\
Three possible scenarios could in principle explain the emission of the NAT and its connection with the relic, and we briefly outline them here:
\begin{itemize} 
    \item (i) NGC4789 is in the pre-shock region, and the shock wave responsible for the relic is moving towards the southwest and approaching the tail. In this scenario, the emission from the tail 
    in the Relic-NAT connection region would be unaffected by the shock passage. However, the plasma injected in the ICM by the tail would play an important role to explain the relic emission, as it would furnish energetic electrons that the low Mach number shock wave would reaccelerate.
    \item (ii) The relic is powered by a quasi-stationary accretion shock, and that NGC4789 is moving supersonically and crossing the shock region from NE to SW. Hence, NGC4789 would be in the pre-shock region, but its tail would have crossed the shock wave responsible for the relic emission. 
    \item (iii) The relic is powered by a shock wave that is moving towards the cluster centre. NGC4789 has been crossed by the shock and it is now in the post-shock region.
\end{itemize}

Scenario (iii) has been recently studied by \cite{Nolting19} using
numerical simulations. They have analysed the case of a shock wave that crosses an AGN when the shock normal is perpendicular to the jets. They find that, when a shock passes through the active jets and lobes of an AGN, jets are distorted by the shock passage, creating a NAT morphology, and a structure similar to the radio bar, as observed here. The spectral index trend resulting from the simulations by \citet{Nolting19} is in agreement with the one observed here along the tail and relic-NAT connection, though the values are slightly flatter after 200 Myr from the shock passage, when the simulation stops. This scenario would require a shock wave moving towards the cluster centre, that seems disfavoured by some authors \citep{Feretti05, Akamatsu13,Churazov21} and also by the spectral index trend across the relic that we detect, which
is steepening NE, in agreement with expectation from an outwards moving shock wave. 
In scenario (ii), the situation would have some similarities with the simulation by \cite{Nolting19}, as also in this scenario the AGN would be crossed by a shock.  However, in this case, the AGN would be in the pre-shock region, and  its tail would be interacting with a less dense environment than the one simulated by \cite{Nolting19}.
Scenario (ii) seems also supported by the analysis of \citet{Adami05}, where they detect a relative velocity of $\sim$1000 km/s between NGC~4789 and NGC~4839, at the cluster centre.
Our analysis does not allow us to discard scenario (i), where the shock is overtaking the pre-existing tail from the back. We can see that the tail and shock must be interacting, because the tail abruptly stops at the position of the relic. This situation has not been simulated yet, but it is likely that it would require some fine-tuning of the parameters.
We indicate scenario (ii) as the favoured one, given the data we have now, though it remains to be seen whether the interaction of the AGN tail with a pre-shocked environment at the cluster outskirts would create the substructures that we observe. 
In all scenarios,  the NAT NGC~4789 would be providing seed electrons that are (re)accelerated by the shock and originate the radio relic. In the literature, there are a few other cases where a link between an AGN tail and a relic has been established (see \citealt{Bonafede14b,vanWeeren17_Nature,Stuardi19}.
\section{Discussion and conclusions}
\label{sec:discussion}
In this work, we have used new data at 144\, MHz from LOFAR to analyse the emission from the Coma cluster. 
We summarise our findings in the following and discuss how these observations allow us to advance our understanding of the non-thermal emission in clusters of galaxies.\\
We have focused our analysis on the properties of the radio halo which - detected at 144\,MHz with the resolution and sensitivity allowed by LOFAR - presents new interesting features and allows us to perform detailed resolved studies of the radio emission. We find that:
\begin{itemize}
\item the radio halo at 144 MHz appears larger than previously reported in the literature, with a largest angular scale of $\sim$71$'$, corresponding to $\sim$ 2 Mpc. The halo is connected to the relic through a low surface brightness radio bridge, and the relic is connected to the AGN NGC\,4879 to the southwest. In total, the radio emission from the halo to the head of NGC4879 spans $\sim$2$^{\circ}$, corresponding to $\sim$3.4 Mpc.
\item The halo brightness profile is well fitted by an exponential elliptical profile. At 144\,MHz, it is characterized by e-folding radii $r_1$= 355 kpc and $r_2$=  268 kpc. At 325 MHz, the profile is more peaked, with $r_1$= 268 kpc and $r_2$=  240 kpc. This is consistent with a spectral steepening of the radio emission towards the halo outskirts, in agreement with the results by \citet{Giovannini93}.
It would be useful to perform these fits also at other frequencies in order to study how the halo size changes with frequency.

\item The spectrum of the radio halo between 144\,MHz and 342\,MHz
is flatter than previously reported, though consistent within the errors. We find $\alpha=-1.0 \pm 0.2$, while previous studies indicate $\alpha \sim - 1.2$. We have computed the flux density of the halo at both frequencies from the best-fit exponential model, that is less affected by the different sensitivities of the two images and the possible presence of residuals from unrelated sources. 

\item We have computed the radial trend of the spectral index $\alpha$ dividing the halo in four symmetric sectors. Our estimates could still be affected by residual contamination from unrelated source and calibration errors, but these should not have a major impact on the global results.
We detect for the first time a moderate steepening towards the cluster center of the spectral index, and confirm the steepening towards the cluster outskirt found at higher frequencies by \citet{Giovannini93}. This trend is in agreement with the expectations of homogeneous turbulent re-acceleration models. Though a detailed modeling is needed to understand the effect of projection affects and the exact location of the steepening frequency, we argue that the steepening detected at the cluster outskirts could indicate a non-constant acceleration time, and hint to a constant turbulent Mach number.
The spectral index steepening towards the cluster centre is more or less pronounced in the different sectors, being prominent in the SW sector and not clearly detected in the NE sector. It is possible that the SW sector has been perturbed by the passage of the NGC4839 group, and shows now different properties.
\item The point-to-point analysis between radio and X-ray surface brightness indicates a sub-linear slope of non-thermal plasma with respect to the thermal plasma. We obtain $I_R\propto I_X^{0.64}$ when the correlation is computed on images at 1$^{\prime}$ resolution. We detect a steeper, yet sublinear, correlation, when the radio image is convolved with Gaussian kernels of 2, 3, 4, 5 and 6 arcmin, and it converges to $I_R \propto I_X^{0.76}$. Indeed, images at lower resolution are more sensitive to the weak emission in the halo outermost regions. We note that the total halo flux density does not change when computed from the 35\asec or 6$^{\prime}$ image, because the outermost regions of the halo yield a very minor contribution to the total halo flux density. However, these regions affect the radio- X-ray correlation making it steeper.
\item We have investigated whether the radio-X-ray correlation has a different slope in the halo core than in the outer halo, finding that the correlation is flatter in the core ($I_R\propto I_X^{0.41}$) than in the outer halo  ($I_R\propto I_X^{0.76}$). By investigating the radial trend of
the quantity $I_R-I_X$, we have confirmed that this trend can be interpreted as a radial trend of $I_R$ versus $I_X$ being flatter in the halo core.
An opposite slope-change has been recently found in some cool-core clusters, where the mini halo emission is surrounded by a weaker and more extended component \cite{Biava21b}. A flatter slope in the halo core is inconsistent with a major contribution of secondary electrons produced through hadronic interaction between thermal protons and cosmic ray protons in the ICM.
\item In the framework of homogeneous re-acceleration models, the change of the slope of the $I_R-I_X$ correlation can be only partially accounted for by a declining magnetic field profile. We have investigated the role of $X$, i.e. the ratio of CRe to thermal energy density, and find that a radially increasing value of $X$ would provide a better match with data. Although more detailed modeling should be done to derive firm conclusions, we note that an increasing value of $X$ with the distance from the cluster centre is also expected from a theoretical point of view. 
\item With the help of MHD cosmological simulations, we have computed the turbulent energy flux associated to the compressive and solenoidal velocity components in a Coma-like cluster, and we find that both quantities show a sub-linear scaling with the simulated X-ray emission, that is in agreement with the observed scaling of $I_R$ versus $I_X$.
Assuming the same efficiency for both modes, the flux associated to the solenoidal velocity component is a factor 10 higher than the flux associated to the compressive component. Hence, once the CRe distribution throughout the volume is assumed, it would be possible to constrain the relative importance of the two modes in the process of particle acceleration.

\end{itemize}

From the analysis of the Coma field, we also conclude that:
\begin{itemize}
    \item To the northeast of Coma, at a projected distance of $\sim$ 1.2 $R_{vir}$, an arc-like diffuse patch of emission is discovered. As a large-scale filament of galaxies is detected in that direction, we tentatively propose that this emission is due to particles re-accelerated by an accretion shock, and name this emission ``accretion relic". If confirmed, this would be the first detection of particle acceleration from an accretion shock.
    \item The halo front, already reported by \citet{BrownRudnick11} is here confirmed coincident with the position of a shock front detected in both X-ray and SZ studies \citep{PlanckComa,Churazov21}. The radio spectral index does not seem to follow the typical trend found in radio relics. However, the large errors due to the small frequency range used to compute the spectral index do not allow us to exclude such a trend. It is possible that the halo front is caused by the radio halo plasma dragged along by the shock wave amd compressed by it. In any case, we can conclude that the properties of the halo in the W region are affected by the shock passage.
    \item The radio relic in the Coma cluster is here imaged with unprecedented detail, thanks to both the sensitivity of the LOFAR observations and to the calibration techniques that we have used, that allow us to minimise the artefacts from Coma A.
    The relic emission is connected to the tail of NGC4879, that is likely moving towards southwest. The connection between the relic and NGC4879 is what we name relic-NAT connection. The 20\asec resolution image shows that the endpoints of the NGC4879 jets are well distinct from the weak diffuse emission of the relic-NAT connection. A bright bar of radio emission is detected, similar to what has been found in other cluster tails \cite[e.g][]{Clarke13,Wilber17}. We have discussed three possible scenarios to explain the presence of the relic-NAT connection, and propose that NGC4879 is moving supersonically towards south-west. During its motion it has crossed the shock at the position of the relic. The shock has re-energised the particles injected by the tail in the ICM in the past and left behind during the galaxy's motion through the ICM.
\end{itemize}

Using literature information about the merging scenario of Coma and its large-scale structure environment, we can outline a global picture to explain the observed radio emission. The Coma cluster is currently accreting matter through filaments of galaxies that connect it to Abell 1367 \citep{Malavasi20}. A recent merger has happened between Coma and the galaxy group NGC~4839, that has passed the cluster from North-East to South-West, injecting a first shock wave in the ICM that is now powering the radio relic emission \citep{Lyskova19}. The cluster core has been perturbed by this merger, and possibly from previous less massive mergers, that have released thermal energy in the ICM. A small fraction of this energy has been dissipated in turbulent motions, that have re-accelerated a mildly relativistic population of CRe already present in the ICM originating the radio halo.\\ 
From the global spectral index of the halo, its radial trend, and the analysis of the radio-X ray correlation, we are able to derive a coherent picture where particles are re-accelerated by homogeneous turbulence in the ICM.
In this picture, we have made several working assumptions that indicate possible regimes for re-acceleration to operate (i.e. constant turbulent Mach number and a radial increase of the CRe energy density with respect to the thermal energy density).
We have attempted to distinguish between re-acceleration by Transit-Time-Damping with compressive modes and non-resonant second order Fermi acceleration with solenoidal modes, and although present data are not accurate enough, we have shown that observations are entering a regime where the details of the model can be in principle tested. It is possible that future observations, either with LOFAR 2.0 and/or with MeerKAT will be able to make an additional step forward and unveil the details of turbulent re-acceleration.\\
The cluster core has been perturbed by the passage of NGC4839, and its motion around its equilibrium position has caused a second shock wave \citep{Lyskova19,Churazov21} whose front is now coincident with the halo front. We find that the spectrum of the halo front has been affected by the shock passage, but we are not able to distinguish between shock re-acceleration or compression by the shock wave.\\ NGC4839 is now at its second passage towards the cluster centre \citep{Lyskova19,Churazov21}. Its motion might have injected additional turbulence in the ICM and a considerable amount of seed electrons, that originate to the radio bridge \citep{Bonafede21}. The NAT NGC~4789 that is now located to the South-West of the relic is moving away from the cluster center at a supersonic velocity after crossing the shock wave at the location of the relic. From this interaction, the radio plasma injected in the ICM by NGC~4789 has been re-energised, leading to the emission that we detect in the relic-NAT connection.\\
Finally, a filament of matter is detected to the Northeast of the Coma cluster. The matter accreting towards the cluster from that direction could result in the tentative ``accretion relic" that we have discovered. \\
The scenario that we outline here is not the only possible one, and it leaves open questions. However, our analysis shows that we are entering a new era for the physics of non-thermal ICM emission, where we can constrain the model parameters.

\section*{Acknowledgements}

AB, EB, NB, CJR  acknowledge support from the ERC Starting Grant `DRANOEL', number 714245. 
AB and CS acknowledge support from the MIUR grant FARE `SMS'
FV, KR, and MBrienza acknowledge support from the ERC Starting Grant `MAGCOW', number 714196.
HB and PM acknowledge financial contribution from the contracts ASI-INAF Athena 2019- 27-HH.0, ``Attivit\'a  di Studio per la comunit\`a  scientifica di Astrofisica delle Alte Energie e Fisica Astroparticellare" (Accordo Attuativo ASI-INAF n. 2017-14-H.0), from the European Union’s Horizon 2020 Programme under the AHEAD2020 project (grant agreement n. 871158) and support from INFN through the InDark initiative. 
MBr\"uggen acknowledges funding by the Deutsche Forschungsgemeinschaft (DFG, German Research Foundation) under Germany’s Excellence Strategy – EXC 2121 ‘Quantum Universe’ – 390833306.
ABotteon acknowledges support from the VIDI research programme with project number 639.042.729, which is financed by the Netherlands Organisation for Scientific Research (NWO). 
RJvW acknowledges support from the ERC Starting Grant ClusterWeb 804208. 
XZ acknowledges support from Chinese Scholarship Council (CSC). 
AS is supported by the Women In Science Excel (WISE) programme of the NWO. 
MR and FG acknowledge support from INAF mainstream project "Galaxy Clusters science with LOFAR". 
LOFAR, the Low Frequency Array designed and constructed by ASTRON, has facilities owned by various parties (each with their own funding sources), and that are collectively operated by the International LOFAR Telescope (ILT) foundation under a joint scientific policy. The LOFAR software and dedicated reduction packages on https://github.com/apmechev/GRID\_LRT were deployed on the e-infrastructure by the LOFAR e-infragroup, consisting of J.B.R.O. (ASTRON \& Leiden Observatory), A.P.M. (Leiden Observatory) and T.S. (Leiden Observatory) with support from N. Danezi (SURFsara) and C. Schrijvers (SURFsara). This research had made use of the NASA/IPAC Extragalactic Database (NED), which is operated by the Jet Propulsion Laboratory, California Institute of Technology, under contract with the National Aeronautics and Space Administration. This research made use of APLpy, an open-source plotting package for Python (Robitaille and Bressert, 2012). 
This research made use of Astropy,\footnote{http://www.astropy.org} a community-developed core Python package for Astronomy \citep{astropy:2013, astropy:2018}



\bibliographystyle{aasjournal}
\bibliography{master}

\appendix
\section{XMM-Newton ObsIDs}\label{appendix:obsid}
The ObsIDs of the \emph{XMM-Newton} observations we analyzed are 58940701, 124710101, 124710301, 124710401, 124710501, 124710601, 124710701, 124710801, 124710901, 124711101, 124711401, 124711601, 124712001, 124712101, 124712201, 124712401, 124712501, 153750101, 204040101, 204040201, 204040301, 300530101, 300530201, 300530301, 300530401, 300530501, 300530601, 300530701, 304320201, 304320301, 304320801, 403150101, 403150201, 403150301, 403150401, 652310201, 652310301, 652310401, 652310501, 652310601, 652310701, 652310801, 652310901, 652311001, 691610201, 691610301, 800580101, 800580201, 851180501, 841680101, 841680201, 841680401, 841680501, 841680301, 841680801, 841680601, and 841680701. 



\end{document}